\documentclass{aa}  

\usepackage{graphicx}
\usepackage{txfonts}
\usepackage{ae,aecompl}
\usepackage{longtable}
\usepackage{graphicx}   
\usepackage{amsmath}    
\usepackage{amssymb}    
\usepackage{dirtytalk}
\usepackage[normalem]{ulem}
\usepackage[table]{xcolor}
\usepackage{siunitx}
\usepackage{natbib}
\bibpunct{(}{)}{;}{a}{}{,} 
\begin{document}

   \title{{\it NuSTAR} spectral analysis of three Seyfert galaxies: NGC 3227, NGC 5548 and MR 2251$-$178}

   \author{Indrani Pal,
          \inst{1}
          C. S. Stalin,\inst{1}
          L. Mallick\inst{3,1}
          and
          Priyanka Rani\inst{2}}

   \institute{Indian Institute of Astrophysics, Bangalore, India\\
              \email{indrani.pal@iiap.res.in}
              \and
             Inter University Centre for Astronomy and Astrophysics, Pune, India\\
             \and
             Cahill Center for Astronomy and Astrophysics, California Institute of Technology, Pasadena, CA 91125, USA\\
             }

   \date{Received October 15, 2021; accepted February 28, 2022}

\abstract
  {The observed nuclear X-ray emission in the radio-quiet category of active galactic nuclei (AGN) is believed to be from a compact region, the corona situated in the vicinity of the central supermassive black holes (SMBH). The shape of the X-ray continuum, among other factors, depends on the temperature of the corona ($\rm{kT_{e}}$). The launch of the {\it Nuclear 
Spectroscopic Telescope Array} ({\it NuSTAR}) has led to the determination of the high energy cut-off ($\rm{E_{cut}}$; and thereby $\rm{kT_{e}}$) in many AGN. In a handful of sources, multiple observations with {\it NuSTAR} have also revealed changes in $\rm{E_{cut}}$. } 
   {In this work we aimed to investigate the variation in $\rm{kT_{e}}$ in three AGN, namely NGC 3227, NGC 5548 and MR 2251$-$178 using more than one epoch of data on a source from {\it NuSTAR}.}
    {We carried out spectral analysis of multiple epochs of data acquired using {\it NuSTAR} on the three sources including a few new observations not published so far. By fitting Comptonization model to the data we determined the temperature of the corona and also investigated changes in $\rm{kT_{e}}$ if any in these sources.}
    {In NGC 3227, we found evidence for variation in $\rm{kT_{e}}$.  We found no correlation of $\rm{kT_{e}}$,   
photon index ($\Gamma$), reflection fraction ($R$) and optical depth ($\tau$) with flux, while, $\tau$ is found to anti-correlate with $\rm{kT_{e}}$. These could be due to more than one physical process at work in the source that causes the change in $\rm{kT_{e}}$. Conclusive evidence for the variation in $\rm{kT_{e}}$ is not found in MR 2251$-$178 and NGC 5548.}
       {}

    
\titlerunning{{\it NuSTAR} spectral analysis of three Seyfert galaxies}
\authorrunning{Indrani et al.} 
\maketitle



\section{Introduction} \label{sec:intro}
Active galactic nuclei (AGN) are amongst the most luminous objects
(L = 10$^{42}$ $-$ 10$^{46}$ erg s$^{-1}$; \citealt{1999PNAS...96.4749F}) 
in the Universe that emit radiation over a wide range of wavelengths. They are believed to be powered by 
accretion of matter onto SMBH (10$^5$ $-$ 
10$^9$ $M_{\odot}$) situated at the center of galaxies \citep{1984ARA&A..22..471R,2008ApJ...689..762D}. The SMBH is generally supposed to be surrounded by an optically thick 
and geometrically thin accretion disk \citep{1973A&A....24..337S}. 
The observed X-ray emission from the nuclear region of the radio-quiet 
category of AGN is believed to be produced by inverse Compton process, caused 
by the interaction of the seed ultra-violet (UV) photons from the accretion disk with the 
thermal electrons in a hot ($\sim$10$^{8-9}$ K) region called the corona that is situated close to the 
accretion disk \citep{1991ApJ...380L..51H,1994ApJ...432L..95H}. 
This X-ray continuum gets reprocessed in the accretion disk giving 
rise to the reflection hump at around 15$-$30 keV and also the 
broad FeK$\alpha$ line at 6.4 keV \citep{1991MNRAS.249..352G, 1993MNRAS.262..179M}. Soft excess between 
0.1 $-$ 2 keV is ubiquitously observed in Type I AGN \citep{1998MNRAS.301..179M, 2002MNRAS.331L..35F, 2006MNRAS.365.1067C,2009A&A...495..421B,2020MNRAS.491..532G}, although the physical origin of this component remains highly debated \citep{2019ApJ...871...88G, 2021ApJ...913...13X}, different analyses showed that a two temperature Comptonization process agrees well with such a component either from an observational (e.g. \citealt{2012MNRAS.420.1825J, 2018A&A...609A..42P,2018A&A...611A..59P,2020A&A...640A..99M,2020MNRAS.497.2352M}) or theoretical (e.g. \citealt{2015A&A...580A..77R, 2020A&A...634A..85P, 2020MNRAS.491.3553B, 2020MNRAS.496.4255B}) point of view. Analysis of these spectral features (reflection, FeK$\alpha$ line, soft excess) will help in providing 
strong constraints on the nature of the X-ray emitting region. 
From X-ray reverberation studies \citep{2009Natur.459..540F,2012MNRAS.422..129Z}
AGN corona is believed to be a compact region situated above the 
accretion disk typically within 3$-$10 $R_G$, where $R_G$ is the gravitational 
radius and it is defined as, $R_G = GM_{BH}/c^2$, here M$_{BH}$ is the SMBH 
mass and $G$ is the gravitational constant. However, there are strong debates concerning the geometry of the corona. The lamp post is one such possibility but other models also exist (e.g. \citealt{1994ApJ...432L..95H, 2012MNRAS.420.1848D, 2013A&A...549A..73P}).Also, rapid X-ray flux variability studies
\citep{2005MNRAS.359.1469M}, the observed small time scales of 
X-ray eclipses \citep{2005ApJ...623L..93R,2011MNRAS.410.1027R}
and microlensing studies \citep{2009ApJ...693..174C} point to the small size 
of the X-ray corona of 5$-$10 $R_G$. 

The observed shape of the X-ray continuum 
can be described by a power law with an exponential cut-off ($\rm{E_{cut}}$), 
and the spectral shape depends on the optical depth ($\tau$), 
temperature of the coronal plasma ($\rm{kT_{e}}$), seed 
photon temperature and the viewing angle. From a study of the Seyfert galaxy NGC 5548, \cite{2000ApJ...540..131P} showed the existence of an approximate relation between $\rm{E_{cut}}$ and $\rm{kT_{e}}$ as $\rm{E_{cut}}$ = 2$-$3 $\rm{kT_{e}}$. On analysis of a sample of Seyfert galaxies, according to \cite{2001ApJ...556..716P}, for an optically thin corona with $\tau$ $<$ 1;  $\rm{E_{cut}}$ $\approx$ 2 $\rm{kT_{e}}$, while for an optically thick corona with $\tau$ $>$ 1,  $\rm{E_{cut}}$ $\approx$ 3 $\rm{kT_{e}}$. However, by fitting Comptonized spectra simulated using a range of $\tau$ and $\rm{kT_{e}}$ with a power law with an exponential cut-off model, \cite{2019A&A...630A.131M}, showed that the commonly adopted relation of $\rm{E_{cut}}$ = 2$-$3 $\rm{kT_{e}}$ is not valid for all values of $\tau$ and $\rm{kT_{e}}$, instead valid for only low values of $\tau$ and $\rm{kT_{e}}$. 

\begin{table*}
\caption{Details of the sources analyzed in this work.  The columns are (1) name of the source, (2) right 
ascension (h:m:s), (3) declination (d:m:s), (4) redshift, (5) galactic hydrogen column density $\rm{N_H}$ in units of $10^{22}$ atoms $\rm{cm^{-2}}$ obtained from \cite{2013MNRAS.431..394W}, (6) type of the 
source, (7) observation ID, (8) epoch, (9) date of observation, and (10) exposure 
time in sec. The information such as the right ascension, declination and $z$ are from \cite{2010A&A...518A..10V}. The OBSIDs that are analysed for the first time are shown in bold.} \label{table-1}
\centering
\begin{tabular}{lllccclclc}
\hline
Name & $\alpha_{2000}$ & $\delta_{2000}$ & $z$ & $\rm{N_H}$ & Type & OBSID & Epoch & Date & Exposure \\
\hline
NGC 3227         & 10:23:30.60   & $+$19:51:56   & 0.004  & 0.021  & Sy1.5 & 60202002002 & A & 09-11-2016 & 49800 \\
                 &  &            &             &                &       & 60202002004 & B & 25-11-2016 & 42457    \\ 
                 &  &           &             &                &       & 60202002006 & C & 29-11-2016 & 39685     \\
                 &  &          &             &                &       & 60202002008 & D & 01-12-2016 & 41812    \\
                 &  &         &             &                &       & 60202002010 & E & 05-12-2016 & 40887     \\
                 &  &        &             &                &       & 60202002012 & F & 09-12-2016 & 39277   \\
                 &  &       &             &                &       & 60202002014 & G & 21-01-2017 & 47602   \\
                 &  &      &             &               &        &  80502609002 & H & 15-11-2019 & 28782  \\
                 & &      &             &               &         & 80502609004 & I & 05-12-2019 & 27690 \\
NGC 5548         & 14:17:59.53 & $+$25:08:12 & 0.017 & 0.017 & Sy1.5 & 	60002044002  & A & 11-07-2013 &                  24096 \\
                 & & & &  &                                                &     60002044003 & B & 12-07-2013 & 27272 \\
                & & & &   &                                                &     60002044005 & C & 23-07-2013 & 49521 \\
                & & & &   &                                                &     60002044006 & D & 10-09-2013 & 51460 \\
                & & & &   &                                                &     60002044008 & E & 20-12-2013 & 50102  \\
                & & & &   &                                                &     90701601002 & F & 26-01-2021 & 38719 \\
MR 2251$-$178    & 22:54:05.90  & $-$17:34:55 & 0.064 & 0.027    & Sy1.5 & 60102025002 & A & 18-05-2015 & 23112  \\
                 &       &       &             &                &       & 60102025004 & B & 17-06-2015 & 23185 \\
                 &       &       &             &                &       & 60102025006 & C & 10-11-2015 & 20588 \\
                 &       &       &             &                &       & 60102025008 & D & 11-12-2015 & 21707   \\
                 &       &       &             &                &       & 90601637002 & E & 16-12-2020 & 23620   \\
\hline
\end{tabular}
\end{table*}

Observations from high energy X-ray missions such as {\it CGRO} 
\citep{2000ApJ...542..703Z,1997ApJ...482..173J}, 
{\it BeppoSAX} \citep{2000ApJ...536..718N,2007A&A...461.1209D}, 
{\it INTEGRAL} \citep{2014ApJ...782L..25M,2010MNRAS.408.1851L,2016MNRAS.458.2454L,
2011A&A...532A.102R}, {\it Swift}-BAT \citep{2013ApJ...770L..37V,2017ApJS..233...17R};
and {\it Suzaku} \citep{2011ApJ...738...70T} showed that the corona in Seyfert galaxies has a wide range of temperature with $\rm{E_{cut}}$ ranging from 50 $-$ 
500 keV. However, observations from those missions are limited to bright and 
nearby sources.  Thus, it is very clear that many efforts were put to measure 
$\rm{E_{cut}}$ in the X-ray spectra of several AGN. However, a major transformation 
in the studies of the Comptonization spectrum of AGN to determine $\rm{E_{cut}}$ 
from an epoch of observation happened after the launch of 
the {\it Nuclear Spectroscopic Telescope Array} ({\it NuSTAR}; \citealt{2013ApJ...770..103H}) in the year 2012, owing to its broad spectral coverage of 3$-$79 keV and its high sensitivity 
beyond 10 keV.  Since the launch of {\it NuSTAR}, $\rm{E_{cut}}$ values were 
obtained for many AGN \citep{2017MNRAS.467.2566F,2018A&A...614A..37T,2018JApA...39...15R,2018ApJ...856..120R,2019MNRAS.484.5113R,2019ApJ...875L..20L,2020ApJ...905...41B, 2021MNRAS.500.1974R,2021MNRAS.502...80K}. Importantly, in addition to the determination of $\rm{E_{cut}}$ values 
(and thereby constraining $\rm{kT_{e}}$), in less than half-a-dozen 
sources there are also reports of variation in the $\rm{E_{cut}}$ values pointing 
to variations in $\rm{kT_{e}}$. 

For example in seven sources, namely MCG-5-23-16 \citep{2017ApJ...836....2Z}, 3C 382 
\citep{2014ApJ...794...62B}, NGC 4593 \citep{2016MNRAS.463..382U}, 
NGC 5548 \citep{2015A&A...577A..38U}, Mrk 335 \citep{2016MNRAS.456.2722K}, NGC 3227 and SWIFT J2127.4+5654 \citep{2021MNRAS.502...80K} variations in the $\rm{E_{cut}}$ values are available in the literature. Recently, 
from a reanalysis of the {\it NuSTAR} spectra of five sources using a model 
independent approach, \cite{2018ApJ...863...71Z} 
confirmed the $\rm{E_{cut}}$  variation in three of the five sources 
namely 3C 382, NGC 5548 and Mrk 335. Most of these inferences were based on variations in $\rm{E_{cut}}$ 
obtained from phenomenological model fits to the data. However, to know changes in 
$\rm{kT_{e}}$ it is imperative to fit physical models to the data, because, it is known
recently, that the relation $\rm{E_{cut}}$ = 2-3 $\rm{kT_{e}}$ does not always hold true \citep{2019A&A...630A.131M}.
Therefore it is imperative to fit physical model fits to the observed spectra to get $\rm{kT_{e}}$. 
Though $\rm{E_{cut}}$ is known to vary, we do not yet know the causes for its variation. Despite that, it is important to increase the number of sources that show variation in the temperature of the corona. This is now achievable 
owing to the multiple-epochs of observation available on a large number of AGN in 
the {\it NuSTAR} archives\footnote{https://heasarc.gsfc.nasa.gov/cgi-bin/W3Browse/w3browse.pl}. The primary motivation here is, therefore to increase 
the number of AGN that show variation in $\rm{kT_{e}}$. We are in the
process of a careful and systematic investigation of
$\rm{kT_{e}}$ variation in a large number of AGN. 
Here, we present the results from the multiple-epoch spectral analysis of
three AGN, namely NGC 3227, NGC 5548 and MR 2251$-$178. This also includes few new observations not published so far.

NGC 3227,at a redshift of z = 0.004 and powered by a black hole of mass 4.79 $\times$ $10^{6}$ $M_{\odot}$ \citep{2015PASP..127...67B}, has been extensively studied in the X-ray band. Signatures of warm absorbers are evident in this source from observations with {\it ASCA} 
\citep{1994ApJ...435..106N,1998ApJ...509..146G}, {\it ROSAT} \citep{1997A&A...327..483K} 
and {\it XMM-Newton} \citep{2009ApJ...691..922M}. In the {\it XMM-Newton} observations, FeK$\alpha$ line 
was evident \citep{2009ApJ...691..922M}. It has also been recently
studied by \cite{2020MNRAS.494.5056L} for flux variations
combining  {\it XMM-Newton} and {\it NuSTAR} observations. It has complex
absorption features, which are also variable \citep{2018MNRAS.481.2470T}. Recently \cite{2021A&A...652A.150M} has reported the broad band spectral modeling of the source using multi-wavelength data from {\it XMM-Newton}, {\it NuSTAR} and {\it Hubble Space Telescope} {\it (HST)}. 
NGC 5548 is a galaxy located
at $z$ = 0.017 with a black hole of mass 
5.0 $\times $ 10$^7$ $M_{\odot}$ \citep{2015PASP..127...67B}. It has been
extensively studied in the X-ray band using data from various satellites
and has also been found to be strongly absorbed in 
soft X-rays \citep{2016A&A...592A..27C, 2015A&A...577A..38U, 2015A&A...575A..22M,2014Sci...345...64K}.
MR 2251$-$178, with a black hole mass of 2.0 $\times $ 10$^8$ $M_{\odot}$ \citep{2009AJ....137.3388W}, first discovered by its strong X-ray emission 
\citep{1978Natur.271...35R} was found to be a low redshift AGN at 
$z$ = 0.06 \citep{1983MNRAS.202..125B}.
From VLA observations, \cite{1990ApJ...356..389M} 
found the source to show weak radio emission
with elongated morphology resembling a FRI source.
Details of the observations and 
reduction of the data are presented in Section 2, analysis is presented 
in Section 3, results and discussion are presented in Section 4
followed by the summary in the final section.

\section{Observations and Data Reduction}
\begin{table*}
\caption{Results of the fit to the spectra of NGC 3227. The models are Model I: {\it const*TBabs*zTBabs(zpo+zgauss)}, Model II: {\it const*TBabs*zTBabs(pexrav+zgauss)}, Model III: {\it const*TBabs*zTBabs(xillver)} and Model IV: {\it const*TBabs*zTBabs(xillverCP)}. The {\it zTBabs} component is added with all the models to fit the epoch G, H and I spectra. The fluxes are in units of 10$^{-10}$ erg cm$^{-2}$ s$^{-1}$ in the 4$-$60 keV band except in epoch I where the flux was calculated in 4$-$50 keV band. Energy (E) of the FeK$\alpha$ line, equivalent width (EW) of the line, $\rm{E_{cut}}$ and $\rm{kT_{e}}$ are expressed in units of keV, model normalization is in units of 10$^{-4}$ photons keV$^{-1}$ cm$^{-2}$s$^{-1}$ and $\rm{N_H}$(zTBabs) is the host galaxy hydrogen column density in units of $10^{22}$ atoms $\rm{cm^{-2}}$. $\rm{C_{FPMA/FPMB}}$ is the cross-calibration constant. The width of the FeK$\alpha$ line was fixed to 0.1 keV during the fitting.}\label{table-2}
\centering
\begin{tabular}{p{0.09\linewidth}p{0.06\linewidth}p{0.06\linewidth}p{0.06\linewidth}p{0.06\linewidth}p{0.06\linewidth}p{0.06\linewidth}p{0.06\linewidth}p{0.06\linewidth}p{0.06\linewidth}}
\hline
Parameter & epoch A & epoch B & epoch C & epoch D & epoch E & epoch F & epoch G & epoch H & epoch I \\
\hline\hline
\multicolumn{10}{c}{Model I : {\it const*TBabs*zTBabs(zpo+zgauss)}} \\
\hline
$\Gamma$ & 1.56$^{+0.01}_{-0.01}$ & 1.55$^{+0.01}_{-0.01}$ & 1.61$^{+0.01}_{-0.01}$ & 1.63$^{+0.01}_{-0.01}$ &  1.66$^{+0.01}_{-0.01}$ & 1.63$^{+0.01}_{-0.01}$ & 1.61$^{+0.02}_{-0.02}$ & 1.64$^{+0.04}_{-0.04}$ & 1.48$^{+0.07}_{-0.07}$   \\
$\rm{N_H}$(zTBabs) & - & - & - & - & -& - & 2.91$^{+1.04}_{-1.03}$ & 4.66$^{+1.83}_{-1.80}$ & 4.57$^{+3.73}_{-3.62}$ \\
E & 6.35$^{+0.03}_{-0.03}$ & 6.38$^{+0.03}_{-0.04}$ & 6.33$^{+0.04}_{-0.04}$ & 6.32$^{+0.04}_{-0.03}$ & 6.30$^{+0.05}_{-0.05}$ & 6.34$^{+0.05}_{-0.05}$ & 6.23$^{+0.04}_{-0.04}$ & 6.28$^{+0.07}_{-0.07}$ & 6.38$^{+0.05}_{-0.05}$   \\
EW & 149$^{+23}_{-32}$ & 188$^{+32}_{-42}$ & 142$^{+30}_{-35}$ & 129$^{+22}_{-24}$ & 86$^{+26}_{-30}$ &  115$^{+26}_{-38}$ & 115$^{+23}_{-23}$ & 123$^{+44}_{-40}$ & 269$^{+66}_{-81}$    \\
norm & 71$^{+2}_{-2}$ & 58$^{+2}_{-2}$ & 73$^{+2}_{-2}$ & 90$^{+3}_{-3}$ & 101$^{+3}_{-3}$ & 92$^{+3}_{-3}$ & 122$^{+7}_{-6}$ &  76$^{+8}_{-7}$ & 17$^{+4}_{-3}$ \\
$\chi^2/dof$ & 930/826 & 752/706 & 834/704 & 909/777 & 815/790 & 684/717 & 1081/935 & 600/595 & 257/247   \\
$\rm{C_{FPMA/FPMB}}$ & 1.04$^{+0.01}_{-0.01}$ & 1.02$^{+0.01}_{-0.01}$ & 1.01$^{+0.01}_{-0.01}$ & 1.01$^{+0.01}_{-0.01}$ & 1.02$^{+0.01}_{-0.01}$ & 1.00$^{+0.01}_{-0.01}$ & 0.99$^{+0.01}_{-0.01}$ & 1.04$^{+0.01}_{-0.01}$ & 1.06$^{+0.03}_{-0.03}$ \\
\hline
\multicolumn{10}{c}{Model II : {\it const*TBabs*zTBabs(pexrav+zgauss)}} \\
\hline
$\Gamma$ & 1.68$^{+0.05}_{-0.05}$ & 1.64$^{+0.06}_{-0.05}$ & 1.77$^{+0.05}_{-0.05}$ & 1.83$^{+0.04}_{-0.04}$ & 1.88$^{+0.05}_{-0.05}$ & 1.82$^{+0.05}_{-0.05}$ & 1.92$^{+0.01}_{-0.07}$ & 1.85$^{+0.08}_{-0.08}$ & 1.70$^{+0.04}_{-0.11}$  \\
$\rm{N_H}$(zTBabs) & - & - & - & - & -& - & 6.43$^{+0.92}_{-1.42}$ & 7.71$^{+1.93}_{-1.95}$ & 8.52$^{+4.30}_{-5.18}$ \\
$\rm{E_{cut}}$ & 212$^{+140}_{-63}$ & 163$^{+118}_{-50}$ & $>$254 & $>$1147 & $>$411 & $>$406 & $>$571 & $>$775 & $>$126 \\
R & 0.52$^{+0.15}_{-0.13}$ & 0.48$^{+0.18}_{-0.15}$ & 0.53$^{+0.18}_{-0.15}$ & 0.55$^{+0.16}_{-0.14}$ & 0.73$^{+0.19}_{-0.17}$ & 0.55$^{+0.18}_{-0.16}$ & 0.83$^{+0.16}_{-0.15}$ & 0.45$^{+0.20}_{-0.18}$ & 0.38$^{+0.49}_{-0.34}$ \\
E & 6.35$^{+0.03}_{-0.03}$ & 6.38$^{+0.04}_{-0.04}$ & 6.33$^{+0.04}_{-0.04}$ & 6.32$^{+0.04}_{-0.04}$ & 6.31$^{+0.06}_{-0.06}$ & 6.34$^{+0.05}_{-0.05}$ & 6.23$^{+0.04}_{-0.04}$ & 6.28$^{+0.07}_{-0.07}$ & 6.38$^{+0.06}_{-0.06}$  \\
EW & 137$^{+32}_{-20}$ & 177$^{+37}_{-42}$ & 129$^{+26}_{-33}$ & 116$^{+25}_{-23}$ & 74$^{+26}_{-21}$ &  102$^{+32}_{-22}$ & 85$^{+23}_{-19}$ & 102$^{+34}_{-31}$ & 231$^{+82}_{-74}$    \\
norm & 84$^{+6}_{-6}$ & 66$^{+6}_{-5}$ & 91$^{+8}_{-7}$ & 119$^{+8}_{-7}$ & 139$^{+10}_{-10}$ & 120$^{+9}_{-9}$ & 207$^{+19}_{-25}$ & 112$^{+18}_{-16}$ & 25$^{+9}_{-9}$ \\
$\chi^2/dof$ & 856/824 & 711/704 & 784/702 & 831/775 & 714/788 & 625/715 & 920/933 & 577/593 & 253/245 \\
$\rm{C_{FPMA/FPMB}}$ & 1.04$^{+0.01}_{-0.01}$ & 1.02$^{+0.01}_{-0.01}$ & 1.01$^{+0.01}_{-0.01}$ & 1.01$^{+0.01}_{-0.01}$ & 1.02$^{+0.01}_{-0.01}$ & 1.00$^{+0.01}_{-0.01}$ & 0.99$^{+0.01}_{-0.01}$ & 1.04$^{+0.01}_{-0.01}$ & 1.06$^{+0.03}_{-0.03}$ \\
\hline
\multicolumn{10}{c}{Model III : {\it const*TBabs*zTBabs(xillver)}}\\
\hline
$\Gamma$ & 1.69$^{+0.01}_{-0.01}$ & 1.69$^{+0.01}_{-0.01}$ & 1.77$^{+0.02}_{-0.01}$ & 1.83$^{+0.01}_{-0.01}$ & 1.80$^{+0.01}_{-0.01}$ & 1.79$^{+0.01}_{-0.01}$ & 1.77$^{+0.01}_{-0.01}$ & 1.87$^{+0.02}_{-0.02}$ & 1.91$^{+0.03}_{-0.03}$ \\
$\rm{N_H}$(zTBabs) & - & - & - & - & -& - & 4.73$^{+0.57}_{-0.56}$ & 8.58$^{+0.98}_{-1.08}$ & 12.12$^{+1.86}_{-1.82}$ \\
$\rm{E_{cut}}$ & 127$^{+14}_{-12}$ & 92$^{+10}_{-8}$ & 204$^{+46}_{-34}$ & $>$439 & 378$^{+152}_{-91}$ & 326$^{+123}_{-78}$ & 248$^{+67}_{-34}$ & $>$510 & $>$198 \\
R & 0.86$^{+0.12}_{-0.12}$ & 1.09$^{+0.17}_{-0.16}$ & 0.85$^{+0.14}_{-0.13}$ & 0.76$^{+0.14}_{-0.12}$ & 0.61$^{+0.12}_{-0.09}$ & 0.71$^{+0.14}_{-0.11}$ & 0.68$^{+0.08}_{-0.08}$ & 0.64$^{+0.16}_{-0.14}$ & 1.36$^{+0.47}_{-0.40}$ \\
norm & 1.98$^{+0.02}_{-0.03}$ & 1.46$^{+0.02}_{-0.05}$ & 2.01$^{+0.03}_{-0.03}$ & 2.98$^{+0.04}_{-0.04}$ & 2.97$^{+0.03}_{-0.03}$ & 2.75$^{+0.04}_{-0.04}$ & 3.73$^{+0.04}_{-0.04}$ & 2.72$^{+0.03}_{-0.04}$  & 0.68$^{+0.03}_{-0.03}$ \\
$\chi^2/dof$ & 868/826 & 728/706 & 785/704 & 837/777 & 735/790 & 627/717 & 976/935 & 579/595 & 260/247 \\
$\rm{C_{FPMA/FPMB}}$ & 1.04$^{+0.01}_{-0.01}$ & 1.02$^{+0.01}_{-0.01}$ & 1.01$^{+0.01}_{-0.01}$ & 1.01$^{+0.01}_{-0.01}$ & 1.02$^{+0.01}_{-0.01}$ & 1.00$^{+0.01}_{-0.01}$ & 0.99$^{+0.01}_{-0.01}$ & 1.04$^{+0.01}_{-0.01}$ & 1.06$^{+0.03}_{-0.03}$ \\
\hline
\multicolumn{10}{c}{Model IV : {\it const*TBabs*zTBabs(xillverCP)}} \\
\hline
$\Gamma$ & 1.78$^{+0.01}_{-0.01}$ & 1.80$^{+0.01}_{-0.01}$ & 1.83$^{+0.01}_{-0.01}$ & 1.84$^{+0.01}_{-0.01}$ & 1.83$^{+0.01}_{-0.01}$ & 1.83$^{+0.01}_{-0.01}$ & 1.82$^{+0.01}_{-0.01}$ & 1.87$^{+0.02}_{-0.02}$ & 1.91$^{+0.04}_{-0.03}$ \\
$\rm{N_H}$(zTBabs) & - & - & - & - & -& - & 4.97$^{+0.59}_{-0.55}$ & 8.17$^{+0.99}_{-0.82}$ & 12.21$^{+1.73}_{-1.65}$ \\
$\rm{kT_{e}}$ & 33$^{+9}_{-7}$ & 28$^{+9}_{-6}$ & 56$^{+131}_{-18}$ & $>$80 & $>$45 & $>$47 & 50$^{+39}_{-10}$ & $>$85 & $>$36 \\
R & 0.79$^{+0.11}_{-0.11}$ & 0.96$^{+0.18}_{-0.10}$ & 0.83$^{+0.15}_{-0.15}$ & 0.78$^{+0.13}_{-0.12}$ & 0.62$^{+0.11}_{-0.11}$ & 0.71$^{+0.14}_{-0.13}$ & 0.67$^{+0.09}_{-0.10}$ & 0.65$^{+0.16}_{-0.15}$ & 1.28$^{+0.60}_{-0.30}$ \\
norm & 1.94$^{+0.02}_{-0.02}$ & 1.52$^{+0.02}_{-0.02}$ & 1.89$^{+0.03}_{-0.03}$ & 2.56$^{+0.02}_{-0.03}$ & 2.55$^{+0.03}_{-0.05}$ & 2.50$^{+0.04}_{-0.04}$ & 3.32$^{+0.03}_{-0.07}$ & 2.24$^{+0.07}_{-0.05}$ & 0.66$^{+0.02}_{-0.02}$ \\
$\chi^2/dof$ & 881/826 & 746/706 & 790/704 & 838/777 & 733/790 & 629/717 & 970/935 & 581/595 & 261/247 \\
$\rm{C_{FPMA/FPMB}}$ & 1.04$^{+0.01}_{-0.01}$ & 1.02$^{+0.01}_{-0.01}$ & 1.01$^{+0.01}_{-0.01}$ & 1.01$^{+0.01}_{-0.01}$ & 1.02$^{+0.01}_{-0.01}$ & 1.00$^{+0.01}_{-0.01}$ & 0.99$^{+0.01}_{-0.01}$ & 1.04$^{+0.01}_{-0.01}$ & 1.06$^{+0.03}_{-0.03}$ \\
Flux & 1.04$^{+0.02}_{-0.01}$ & 0.86$^{+0.01}_{-0.01}$ & 0.94$^{+0.01}_{-0.02}$ & 1.11$^{+0.01}_{-0.01}$ & 1.15$^{+0.01}_{-0.02}$ & 1.13$^{+0.01}_{-0.01}$ & 1.55$^{+0.01}_{-0.01}$ & 0.88$^{+0.01}_{-0.01}$ & 0.27$^{+0.01}_{-0.01}$ \\
\hline\hline
\end{tabular}
\end{table*}

\subsection{Data reduction}
We reduced {\it NuSTAR} data in the 3$-$79 keV band using the standard {\it NuSTAR} data reduction software
NuSTARDAS\footnote{https://heasarc.gsfc.nasa.gov/docs/nustar/analysis/nustar swguide.pdf} distributed by HEASARC within HEASoft v6.29. Considering the passage of the satellite through the South Atlantic Anomaly (SAA) we selected, SAACALC \say{2}, SAAMODE \say{optimized} and also excluded the tentacle region. The calibrated, cleaned, 
and screened event files were generated by running the {\tt nupipeline} 
task using the CALDB release 20210701. To extract the source counts we chose a circular region of radius 60 arcsec 
centered on the source. Similarly, to extract the background counts,  
we  selected a circular region of the same radius away 
from the source on the same chip to avoid contamination from source photons. We then used the  
{\tt nuproducts} task to generate energy spectra, 
response matrix files (RMFs) and auxiliary response files (ARFs), for both 
the hard X-ray detectors housed inside the corresponding focal plane modules FPMA and FPMB. 
For spectral analysis, using XSPEC version 12.12.0 \citep{1996ASPC..101...17A}, 
we fitted the background subtracted spectra from
FPMA and FPMB simultaneously (without combining them)  allowing the 
cross normalization factor to vary freely during spectral fits. The spectra were binned to have a S/N ratio greater than 5 in each spectral channel using the {\it NuSTAR}-specific Python script {\it snrgrppha}\footnote{https://sites.astro.caltech.edu/$\sim$mislavb/}. To get an estimate of the  model parameters that best describe the observed data, we
used the chi-square ($\chi^2$) statistics and for calculating the errors
in the model parameters we used the $\chi^2$ = 2.71 criterion i.e. 90 per cent confidence
range in XSPEC. 

\begin{table*}
\caption{Results of the model fits to the spectra of MR 2251$-$178. The models are Model I: {\it const*TBabs*zTBabs(zpo)}, Model II: {\it const*TBabs*zTBabs(pexrav)}, Model III: {\it const*TBabs*zTBabs(xillver)} and Model IV: {\it const*TBabs*zTBabs(xillverCP)}. The {\it zTBabs} component is added with all the models to fit the epoch E spectra. In epoch D the width of the Fek$\alpha$ line was fixed to 0.1 keV. In the case where no line was detected, the upper limit on the EW was calculated by fixing the line energy to 6.4 keV.The fluxes are in units of 10$^{-10}$ erg cm$^{-2}$ s$^{-1}$ in the 4$-$60 keV band except in epoch E where the flux was calculated in 4$-$50 keV band. Columns and parameters have the same meaning as given in Table \ref{table-2}.}\label{table-3}
\centering
\begin{tabular}{p{0.12\linewidth}p{0.12\linewidth}p{0.12\linewidth}p{0.12\linewidth}p{0.12\linewidth}p{0.12\linewidth}}
\hline
Parameter & epoch A & epoch B & epoch C & epoch D & epoch E  \\
\hline\hline
\multicolumn{6}{c}{Model I : {\it const*TBabs*zTBabs(zpo+zgauss)}} \\
\hline
$\Gamma$ & 1.75$^{+0.02}_{-0.02}$ & 1.79$^{+0.01}_{-0.01}$ & 1.79$^{+0.02}_{-0.02}$ & 1.79$^{+0.02}_{-0.02}$ & 1.83$^{+0.05}_{-0.05}$ \\
$\rm{N_H}$(zTBabs) & - & - & - & - & 7.82$^{+2.99}_{-2.93}$ \\
E & - & - & - & 6.49$^{+0.37}_{-0.27}$ & - \\
EW & $<$35 & $<$49 & $<$34 & $<$70 & $<$46 \\ 
norm & 142$^{+5}_{-5}$ & 177$^{+6}_{-6}$ & 162$^{+6}_{-6}$ & 162$^{+6}_{-6}$ & 84$^{+13}_{-11}$  \\
$\chi^2/dof$ & 583/601 & 656/633 & 574/551 & 516/576 & 446/400 \\
$\rm{C_{FPMA/FPMB}}$ & 1.02$^{+0.02}_{-0.02}$ & 1.00$^{+0.01}_{-0.01}$ & 1.03$^{+0.02}_{-0.02}$ & 1.02$^{+0.02}_{-0.02}$ &  1.05$^{+0.02}_{-0.02}$  \\
\hline
\multicolumn{6}{c}{Model II : {\it const*TBabs*zTBabs(pexrav+zgauss)}} \\
\hline
$\Gamma$ & 1.65$^{+0.05}_{-0.05}$ & 1.72$^{+0.05}_{-0.03}$ & 1.76$^{+0.07}_{-0.07}$ & 1.79$^{+0.06}_{-0.06}$ & 1.82$^{+0.11}_{-0.10}$  \\
$\rm{N_H}$(zTBabs) & - & - & - & - & 7.66$^{+3.26}_{-3.02}$ \\
$\rm{E_{cut}}$ & 125$^{+96}_{-39}$ & 185$^{+200}_{-69}$ & 110$^{+70}_{-32}$ & 193$^{+417}_{-80}$ & $>$175 \\
R & $<$0.07 & $<$0.11 & 0.29$^{+0.22}_{-0.18}$ & 0.19$^{+0.20}_{-0.17}$ & $<$0.23 \\
E & - & - & - & 6.48$^{+0.67}_{-0.31}$ & - \\
EW & $<$49 & $<$43 & $<$33 & $<$63 & $<$50 \\
norm & 111$^{+8}_{-8}$ & 144$^{+11}_{-6}$ & 139$^{+15}_{-13}$ & 144$^{+15}_{-13}$ & 75$^{+19}_{-14}$\\
$\chi^2/dof$ & 569/599 & 647/631 & 552/549 & 509/574 & 446/398  \\
$\rm{C_{FPMA/FPMB}}$ & 1.02$^{+0.02}_{-0.02}$ & 1.00$^{+0.01}_{-0.01}$ & 1.03$^{+0.02}_{-0.02}$ & 1.02$^{+0.02}_{-0.02}$ &  1.05$^{+0.02}_{-0.02}$  \\
\hline
\multicolumn{6}{c}{Model III : {\it const*TBabs*zTBabs(xillver)}} \\
\hline
$\Gamma$ & 1.65$^{+0.02}_{-0.02}$ & 1.72$^{+0.02}_{-0.02}$ & 1.70$^{+0.02}_{-0.02}$ & 1.77$^{+0.02}_{-0.02}$ &  1.82$^{+0.02}_{-0.03}$ \\
$\rm{N_H}$(zTBabs) & - & - & - & - & 7.56$^{+1.69}_{-1.38}$ \\
$\rm{E_{cut}}$ & 124$^{+22}_{-18}$ & 169$^{+45}_{-30}$ & 103$^{+18}_{-14}$ & 163$^{+46}_{-30}$ & $>$366 \\
R & $<$0.10 & $<$0.16 & 0.17$^{+0.14}_{-0.13}$  & 0.22$^{+0.15}_{-0.13}$ & $<$0.11 \\
norm & 2.92$^{+0.02}_{-0.04}$ & 3.54$^{+0.03}_{-0.05}$ & 2.79$^{+0.05}_{-0.05}$ & 3.05$^{+0.02}_{-0.05}$ & 2.10$^{+0.03}_{-0.52}$ \\
$\chi^2/dof$ & 569/599 & 647/631 & 556/549 & 510/576 & 447/398 \\
$\rm{C_{FPMA/FPMB}}$ & 1.02$^{+0.02}_{-0.02}$ & 1.00$^{+0.01}_{-0.01}$ & 1.03$^{+0.02}_{-0.02}$ & 1.02$^{+0.02}_{-0.02}$ &  1.05$^{+0.02}_{-0.02}$  \\
\hline
\multicolumn{6}{c}{Model IV : {\it const*TBabs*zTBabs(xillverCP)}} \\
\hline
$\Gamma$ & 1.76$^{+0.01}_{-0.01}$ & 1.79$^{+0.01}_{-0.01}$ & 1.80$^{+0.02}_{-0.01}$ & 1.83$^{+0.02}_{-0.01}$ & 1.83$^{+0.02}_{-0.02}$  \\
$\rm{N_H}$(zTBabs) & - & - & - & - & 7.54$^{+1.42}_{-1.45}$ \\
$\rm{kT_{e}}$ & 25$^{+26}_{-6}$ & 35$^{+149}_{-11}$ & 21$^{+8}_{-4}$ & 35$^{+67}_{-11}$ & $>$32  \\
R & $<$0.06 & $<$0.11 & $<$0.25 & 0.17$^{+0.14}_{-0.13}$ & $<$0.09  \\
norm & 2.68$^{+0.02}_{-0.03}$ & 3.26$^{+0.03}_{-0.05}$ & 2.66$^{+0.04}_{-0.04}$ &  2.85$^{+0.04}_{-0.04}$ & 1.71$^{+0.02}_{-0.04}$ \\
$\chi^2/dof$ & 576/599 & 650/631 & 555/549 & 511/576 & 447/398 \\
$\rm{C_{FPMA/FPMB}}$ & 1.02$^{+0.02}_{-0.02}$ & 1.00$^{+0.01}_{-0.01}$ & 1.03$^{+0.02}_{-0.02}$ & 1.02$^{+0.02}_{-0.02}$ &  1.05$^{+0.02}_{-0.02}$  \\
Flux & 1.07$^{+0.01}_{-0.01}$ & 1.22$^{+0.01}_{-0.01}$ & 1.08$^{+0.01}_{-0.01}$ & 1.09$^{+0.01}_{-0.01}$ & 0.47$^{+0.02}_{-0.01}$ \\
\hline\hline
\end{tabular}
\end{table*}

\begin{figure*}
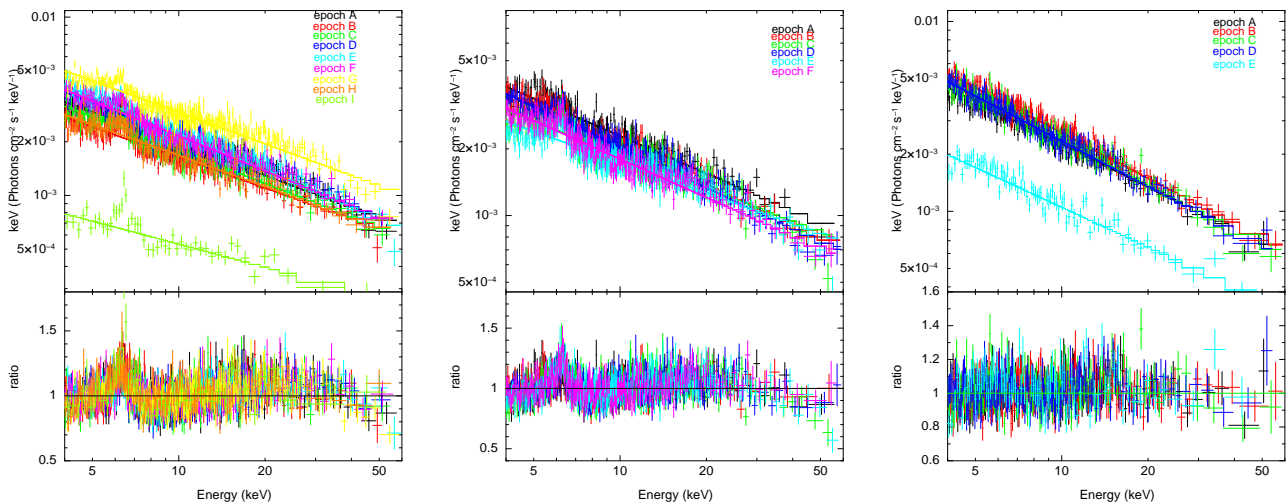

\hbox{
     \includegraphics[scale=0.28]{spectra3227.eps}
     \includegraphics[scale=0.28]{spectra5548.eps}
     \includegraphics[scale=0.28]{spectramr.eps}
     }
\caption{The unfolded spectra of the nine observations for 
	NGC 3227 (left), six observations for NGC 5548 (middle) and five observations 
	for MR 2251$-$178 (right) fitted with a simple power-law. The bottom panels show the 
	ratio of the observed spectra to the model. For clarity, we used only FPMA data. The spectra are rebinned for visualization purpose only. }\label{figure-1}	
\end{figure*}

\begin{figure}
\vbox{
      \includegraphics[scale=0.95]{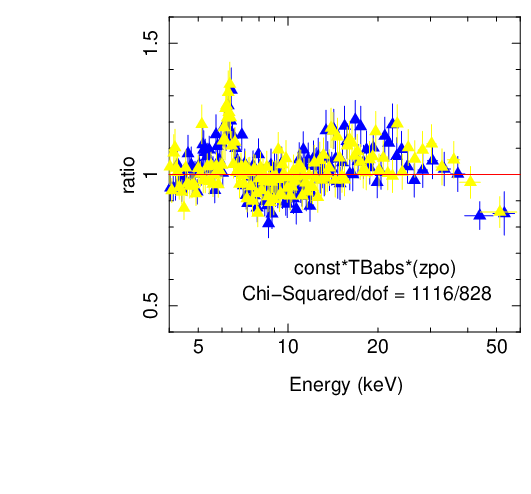}
      \includegraphics[scale=0.95]{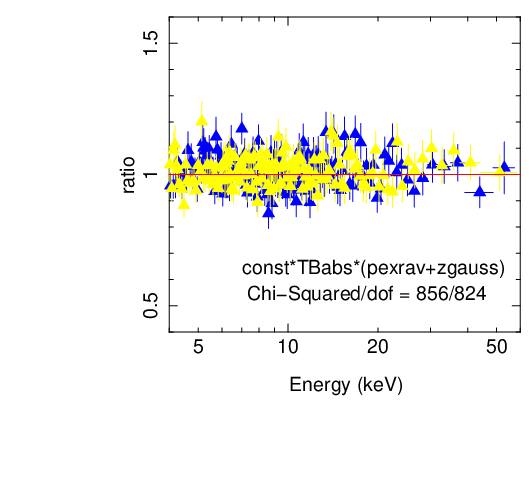}
      \includegraphics[scale=0.95]{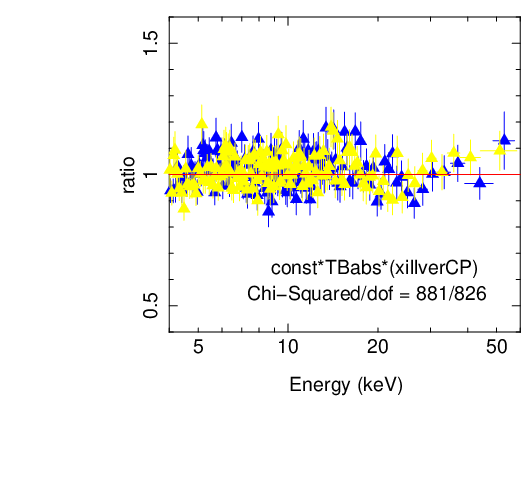}
    }
	\vspace{-1.0cm}
\caption{Ratio of data to the  model for the model fits {\it const*TBabs(zpo+zgauss)},
{\it const*TBabs(pexrav+zgauss)} and {\it const*TBabs(xillverCP)} to the FPMA (blue triangle) and FPMB (yellow triangle) spectra of OBID 60202002002 of NGC 3227. The spectra are rebinned for visualization purpose.}
	\label{figure-2}
\end{figure}

\begin{figure}
\vbox{
      \includegraphics[scale=0.95]{a_ngc5548.eps}
      \includegraphics[scale=0.95]{b_ngc5548.eps}
      \includegraphics[scale=0.95]{c_ngc5548.eps}
    }
\vspace{-1.0cm}
\caption{Ratio plots for the model fits {\it const*TBabs*zTBabs*(zpo+zgauss)}, {\it const*TBabs*zTBabs*(pexrav+zgauss)} and {\it const*TBabs*zTBabs(xillverCP)} to the FPMA (blue star) and FPMB (yellow star) spectra of OBID 60002044006 of NGC 5548. The spectra are rebinned for visualization purpose only.}
\label{figure-3}
\end{figure}

\begin{figure}
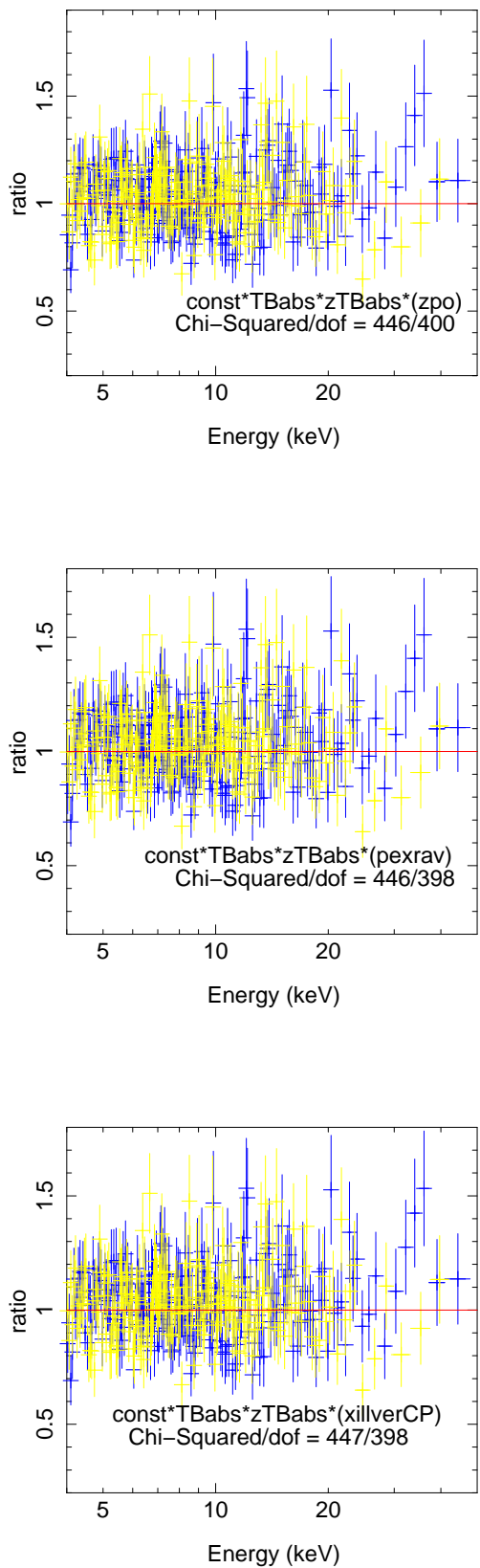

\vbox{
\includegraphics[scale=0.95]{a_mr.eps}
\includegraphics[scale=0.95]{b_mr.eps}
\includegraphics[scale=0.95]{c_mr.eps}
      }
\vspace{-1.0cm}
\caption{Ratio of data to model for the model fits {\it const*TBabs*zTBabs*(zpo)}, {\it const*TBabs*zTBabs*(pexrav)} and {\it const*TBabs*zTBabs(xillverCP)} to the FPMA (blue dot) and FPMB (yellow dot) spectra of OBID 90601637002 of MR 2251$-$178. We re-binned the spectra for visualization purpose only.}
\label{figure-4}
\end{figure}

\section{Analysis of the data}
For few epochs of the sources studied in this work, we do have observations in the soft band from telescopes such as {\it XMM-Newton} for NGC 3227 and {\it XMM-Newton}, {\it Chandra} for NGC 5548 and {\it XMM-Newton} for MR 2251$-$178. However, for this work, we decided to use only {\it NuSTAR} data as (a) the good sensitivity of {\it NuSTAR} over the 3$-$79 keV energy band captures all the key reflection features of an AGN spectrum and as the main goal of this work is to model the high energy rollover of the Comptonized spectra we did not want the absorption in the soft band affecting our analysis in the determination of $\rm{kT_{e}}$ and (b) observations in the soft band are not available simultaneous to the {\it NuSTAR} observations for all the epochs and for all the sources. However we note here, the inclusion of soft X-ray data in the fitting might have an effect on the photon index ($\Gamma$) obtained from using the {\it NuSTAR} data alone. However, a simplest approach to constrain $\rm{kT_{e}}$ (which is the aim of this work) is the use of only {\it NuSTAR} data, but for better constraining the other physical characteristics of the sources broad band spectral analysis including data from UV to hard X-ray band is more appropriate. While analyzing only {\it NuSTAR} data, we ignored the 3$-$4 keV band to limit the effect of absorption if any and also did not consider data in the energy range beyond 60 keV due to lack of source photons. Thus, we carried out spectral fits to the {\it NuSTAR} data in the 4$-$60 keV energy band for all the OBSIDS except for epoch I of NGC 3227, epoch F of NGC 5548 and epoch E of MR 2251$-$178. Due to the unavailability of photons beyond 50 keV we restricted the spectral fit in the 4$-$50 keV energy band to the epoch I and epoch E spectra of NGC 3227 and MR 2251$-$178 respectively. Similarly, for the epoch F spectrum of NGC 5548, we used the FPMA/FPMB data in the 4$-$55 keV range. (see Fig \ref{figure-1}).
\subsection{Phenomenological spectral fits}
For our spectral fits, to model the line of sight galactic absorption, the value of the neutral hydrogen column density ($\rm{N_H}$) for all the sources were frozen to the values obtained from \cite{2013MNRAS.431..394W}. These $\rm{N_H}$ values are given in Table \ref{table-1}. Similarly, the redshifts of the sources were frozen to their corresponding values given in Table \ref{table-1}. Also, we used the solar abundances from \cite{2000ApJ...542..914W} and the photoelectric cross sections from \cite{1996ApJ...465..487V}. For models that require inclination angle ($i$) , we used $i$ = 50$^{\circ}$ for NGC 3227 (\citealt{1997ApJ...477..623S}, \citealt{2019A&A...628A..65A}, \citealt{2016MNRAS.457.1568M}), $i$ = 30$^{\circ}$ for NGC 5548 \citep{2015A&A...577A..38U} and $i$ = 60$^{\circ}$ (i.e., the default value) for MR 2251$-$178.
\subsubsection{Absorbed power law}
Firstly, to understand the continuum emission in our sample of sources we 
fitted the observed X-ray spectra with the baseline phenomenological absorbed power law
model  and having the following form in XSPEC
\begin{equation}
const*TBabs(zpo) 
\end{equation}
The first component of this model is the constant used to calibrate the two focal plane modules of {\it NuSTAR}. The second component, {\it TBabs} \citep{2000ApJ...542..914W} was used to model the line 
of sight galactic absorption. The parameters
that were kept free are $\Gamma$ and the normalization (i.e. photons keV$^{-1}$ cm$^{-2}$ s$^{-1}$). We found evidence of intrinsic absorption present in the lower energy end for all the epochs in  NGC 5548, epoch G, H and I in NGC 3227 and epoch E in MR 2251$-$178. We therefore included a {\it zTBabs} component with the absorbed power law model to fit their spectra and the model looks like 
\begin{equation}
const*TBabs*zTBabs(zpo) 
\end{equation}
in XSPEC. For all epochs in NGC 5548, two epochs in NGC 3227 and one epoch of MR 2251$-$178 non-inclusion of {\it zTBabs} returned a poor fit with $\chi^2/dof$ larger than 1.2.  Inclusion of {\it zTBabs} with $\rm{N_H}$({\it zTBabs}) kept free improved the fit with $\chi^2/dof$ close to unity.

\begin{figure}
\vbox{
      \hspace{-1.0cm}\includegraphics[scale=0.40]{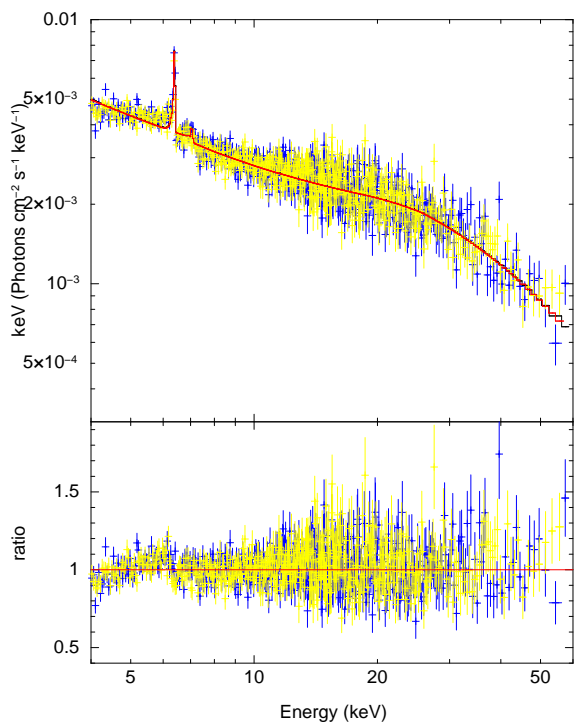}
      }
\caption{The unfolded spectra with the ratio of data to model for the model fits {\it const*TBabs(xillverCP)} to the FPMA (blue) and FPMB (yellow) spectra of OBID 60202002014 (epoch G) of NGC 3227.}
\label{figure-5}		
\end{figure}

From the residual spectra obtained from the simple absorbed power law model fit 
to the observations, we noticed the presence of the fluorescent FeK$\alpha$ line 
in  NGC 3227, NGC 5548 and in epoch D spectra of MR 2251$-$178, but in the other observations of MR 2251$-$178, the residual spectra do not show conspicuous FeK$\alpha$ line. The spectral fits are shown in Fig. \ref{figure-1} for NGC 3227 and NGC 5548 and MR 2251$-$178.
Though this line is common in most of  the
X ray spectra of AGN \citep{1993ARA&A..31..717M, 2007MNRAS.382..194N, 2010HEAD...11.0708D} there are exceptions \citep{2011MNRAS.416..629B}. 
The apparent non-detection of FeK$\alpha$ line in the spectra of MR 2251$-$178 
could be due to weaker reflection owing to larger viewing angle \citep{2011MNRAS.416..629B},
low signal-to-noise ratio spectra, very high ionized accretion disk 
\citep{1993MNRAS.261...74R,1994MNRAS.266..653Z} or a combination of all.
However, the presence of a weak FeK$\alpha$ line in epoch D of MR 2251$-$178 could point to the physical characteristic of MR 2251$-$178 being different from the other two sources.  A thorough analysis is needed to know the exact reasons for the absence/weakness of this line 
but this issue is beyond the scope of this work as we are
in this work mainly interested in the changes in the temperature of the corona. 
To model the FeK$\alpha$ line seen in the residual spectra in NGC 3227, 
NGC 5548 and in one observation (epoch D) of MR 2251$-$178, we included a gaussian component and with this inclusion the model takes the form
{\it const*TBabs(zpo + zgauss)} and the quality of the fit improved. 

After the inclusion of a gaussian component with the power law in the case of NGC 3227, in all nine epochs the $\chi^2$ value reduced in the range between 60 and 242 for a reduction of 2 dof. In epochs C, D, G and I, non-inclusion of the gaussian component, resulted in a reduced $\chi^2$ greater than 1.30. After the inclusion of the gaussian component the reduced $\chi^2$ ranged between 1.04$-$1.16 in these epochs. For the other epochs, the $\chi^2/dof$ was $>$ 1.1 before and it became $\sim$ 1.0 after the line inclusion. For MR2251$-$178, gaussian component was used to fit the FeK$\alpha$ line only in epoch D. Inclusion of the line component lead to a change in $\chi^2$ of 5 for a reduction of 2 dof. This negligible change in $\chi^2$ over 2 dof did not improve the fit quality significantly in this case. For NGC 5548, on inclusion of gaussian line component in all the epochs, the value of the $\chi^2$ reduced in the range 23 to 105 for a reduction of 2 dof. Just an absorbed power law fit to the data produced the reduced $\chi^2$ greater than 1.2 for the epochs C, D, E and F. Adding the Gaussian with the power law lead to a reduced $\chi^2$ of $\sim$ 1.1 for epoch C and D and $\sim$ 1.0 for epoch E and F. Similarly, for epochs A and B, the $\chi^2/dof$ changed from 1.10 to 1.06 and 1.02 to 0.97 respectively. The width of the FeK$\alpha$ line was frozen to the value of 0.1 keV during the fitting, letting it free to vary did not improve the fit significantly. The best fit  
parameters for the sources are given in 
Table  \ref{table-2}, \ref{table-3} and \ref{table-4} for NGC 3227, MR 2251$-$178 and NGC 5548. 
\subsubsection{Pexrav}
In the residuals of the simple power law fit to all the spectra of the sources (see Fig. \ref{figure-1}) we found 
the signature of a high energy turn over and a reflection hump beyond 15 keV. To appropriately model both the high energy cut off and the reflection feature present in the spectra we replaced the {\it zpo} component in our earlier
model with {\it pexrav} and the new model has the form
\begin{equation}
const*TBabs(pexrav).
\end{equation}
While modelling the reflection component of NGC 5548, epoch G, H and I spectra of NGC 3227 and epoch E spectrum of MR 2251$-$178 an intrinsic absorption component, {\it zTBabs} was added with the above model. The intrinsic hydrogen column density, $\rm{N_H}$({\it zTBabs}) was kept free during the fit.
\begin{figure}
\vbox{
     \includegraphics[scale=0.45]{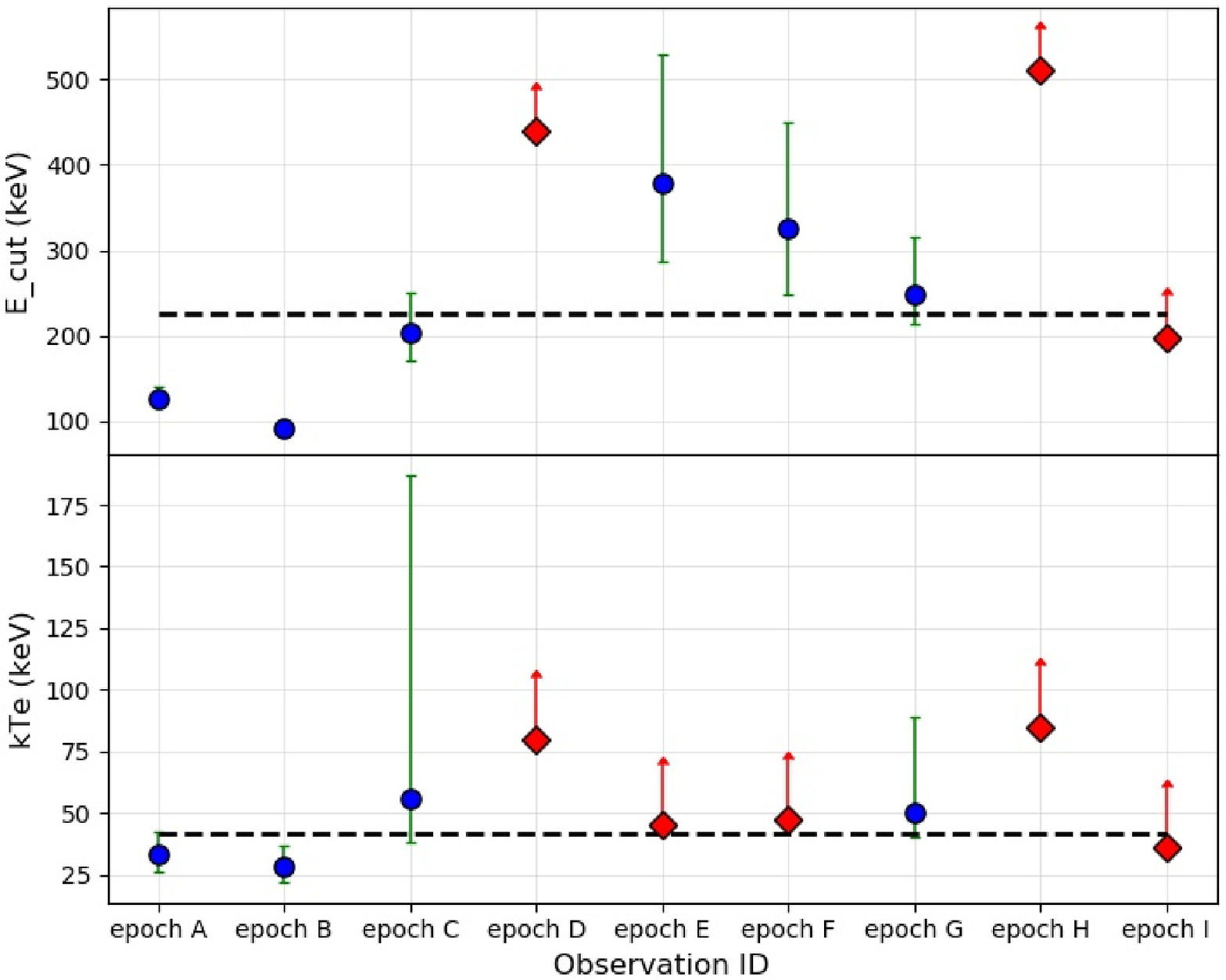}
      }
\caption{Variation of $\rm{E_{cut}}$ and $\rm{kT_{e}}$ with the observation epochs as obtained from {\it xillver} and {\it xillverCP} model fit respectively to the spectra NGC 3227. The plotted errors were calculated using the $\chi^2$ = 2.71 criterion i.e. 90\% confidence range. The black dashed lines in each panel are fits of constant (mean of $\rm{E_{cut}}$ and $\rm{kT_{e}}$) to the data points. For the fitting, epochs in which we were unable to constrain $\rm{E_{cut}}$ and $\rm{kT_{e}}$ were dropped (indicated with red point).} \label{figure-6}		
\end{figure}

\begin{figure*}
\vbox{
     \hspace{-0.2cm}\includegraphics[scale=0.30]{x02_3227.eps}
     \hspace{-0.2cm}\includegraphics[scale=0.30]{x04_3227.eps}
     \hspace{-0.2cm}\includegraphics[scale=0.30]{x06_3227.eps}
     }
\vspace{-0.5cm}
\end{figure*}
\begin{figure*}
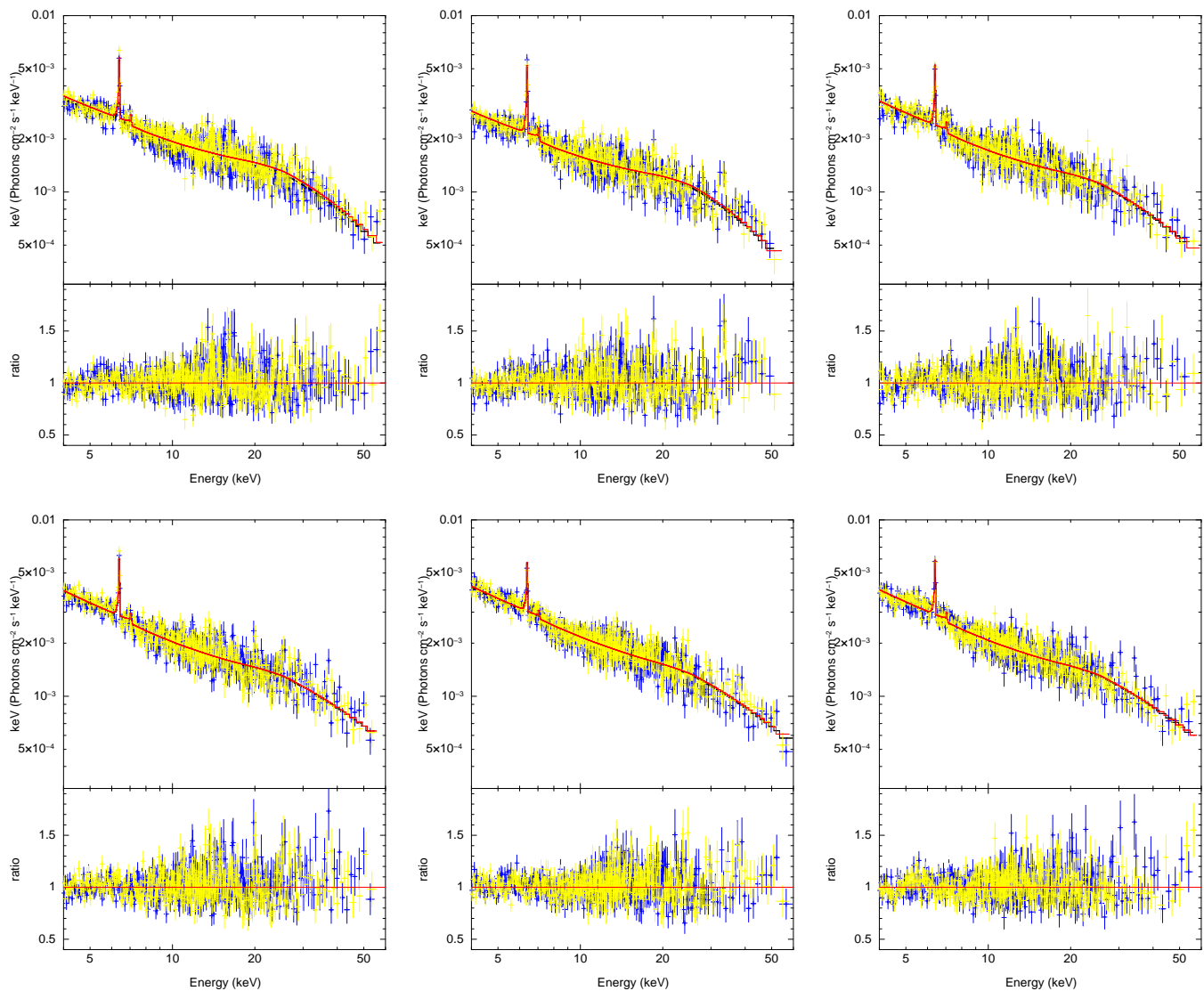

\vbox{
     \hspace{-0.2cm}\includegraphics[scale=0.30]{x08_3227.eps}
     \hspace{-0.2cm}\includegraphics[scale=0.30]{x10_3227.eps}
     \hspace{-0.2cm}\includegraphics[scale=0.30]{x12_3227.eps}
     }
\vspace{-0.5cm}
\caption{The unfolded spectra along with {\it const*TBabs(xillverCP)} model fits and the data to model ratio for the epoch A (top left panel), epoch B (top middle panel), epoch C (top right panel), epoch D (bottom left panel), epoch E (bottom middle panel) and epoch F (bottom right panel) for the source NGC 3227. blue and yellow data points refer to FPMA and FPMB respectively.}
\label{figure-7}
\end{figure*}
\begin{figure*}
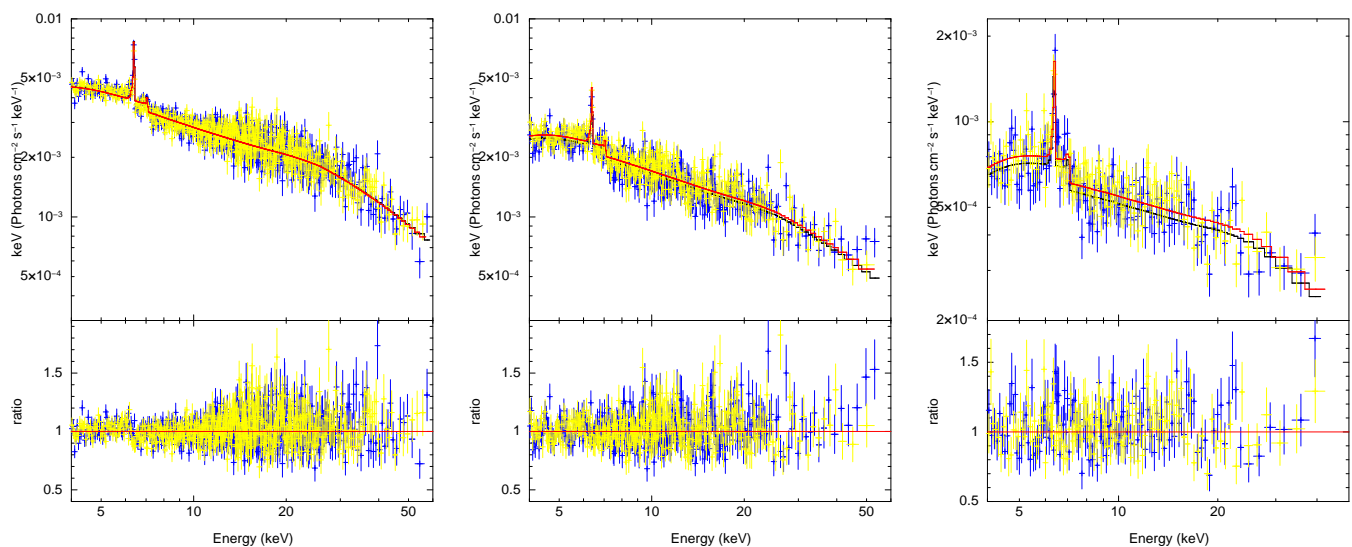

\vbox{
     \hspace{-0.2cm}\includegraphics[scale=0.30]{x14_3227.eps}
     \hspace{-0.2cm}\includegraphics[scale=0.30]{x82_3227.eps}
     \hspace{-0.2cm}\includegraphics[scale=0.30]{x84_3227.eps}
     }
\vspace{-0.5cm}
\caption{The unfolded spectra and model fits along with the data to model ratio for {\it const*TBabs(xillverCP)} fits to the epoch G (left), epoch H (middle) and epoch I (right) of NGC 3227. For epoch G, epoch H and epoch I {\it zTBabs} was added to the model. blue and yellow data points refer to FPMA and FPMB respectively.}
\label{figure-8}
\end{figure*}

This model implements both photoelectric absorption and Compton scattering,  
however, it does not include fluorescence. Therefore to model the FeK$\alpha$ 
line seen in the residual spectra in NGC 3227, NGC 5548 and epoch D in MR 2251$-$178, we included a gaussian component. So, the model takes the following form,
\begin{equation}
const*TBabs(pexrav+zgauss)
\end{equation}
in XSPEC. The model {\it pexrav} improved the fitting compared to {\it zpo} as evident in 
the reduced $\chi^2$ in Tables \ref{table-2}, \ref{table-3} and 
\ref{table-4}. This model includes primary emission 
having a power law form with an exponential cut-off and a reflection component. 
The reflector is considered to be an optically thick medium in an infinite plane 
geometry and covering a larger fraction of the X-ray source. A parameter that comes as an output in the model fit is the reflection parameter $R$. This parameter gives a measure of the reflection component
present in the observed spectra of the sources. For an isotropic source, this
parameter is related to the solid angle ($\Omega$) as $R \sim \Omega/2\pi$, and it is
dependent on the angle of inclination ($i$) between the perpendicular
to the accretion disk and the line of sight to the observer. The width of the FeK$\alpha$ line was fixed at 0.1 keV, treating the parameter free did not improve the fit. For all the sources, the parameters that were left free in the model fits are $\rm{E_{cut}}$, $\Gamma$, R and normalization.

Using {\it (pexrav+zgauss)} in all the nine epochs of NGC 3227, the $\chi^2$ reduced in the range between 4 and 161 for a reduction of 2 dof compared to the {\it (zpo+zgauss)} fit. In MR 2251$-$178, the reduction in $\chi^2$ is in the range between 0 and 22 with a reduction of 2 in dof. For all the six epochs in NGC 5548, the {\it (pexrav+zgauss)} fit produced a reduction in the $\chi^2$ values between 9$-$60 with a reduction of 2 dof compared to (zpo+zgauss). The best fit parameters are given 
in Tables \ref{table-2} , \ref{table-3} and \ref{table-4}. 

\begin{figure*}
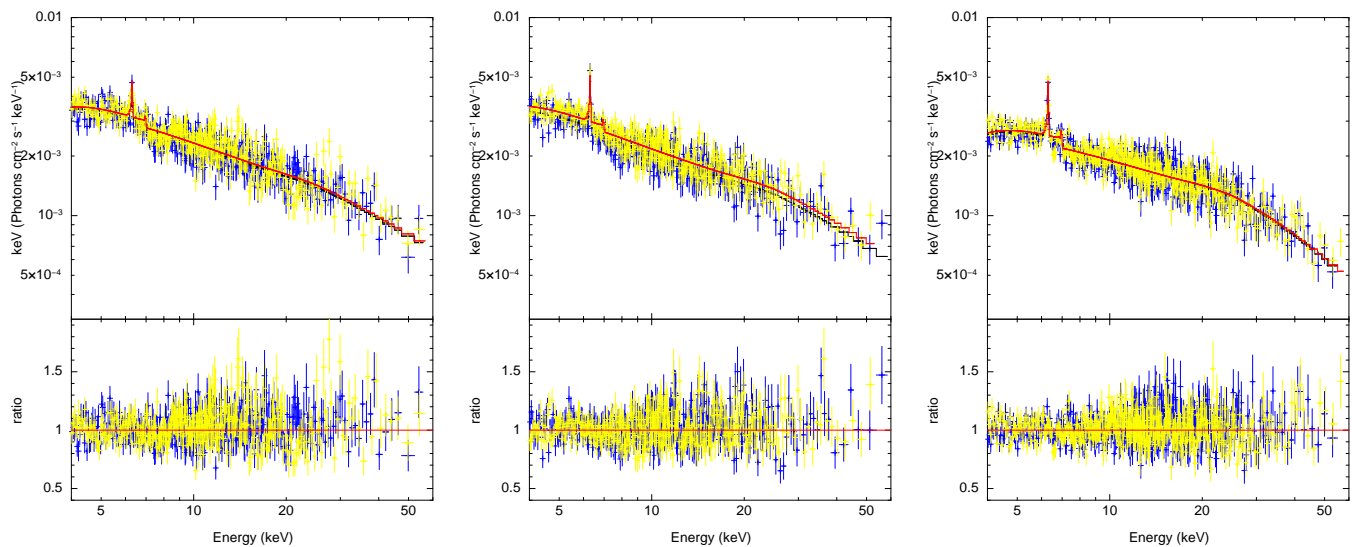

\vbox{
     \hspace{-0.2cm}\includegraphics[scale=0.30]{x02_5548.eps}
     \hspace{-0.2cm}\includegraphics[scale=0.30]{x03_5548.eps}
     \hspace{-0.2cm}\includegraphics[scale=0.30]{x05_5548.eps}
     }
     \vspace{-0.5cm}
\caption{The unfolded spectra with the {\it const*TBabs*zTBabs(xillverCP)} model fits and the data to model ratio to the spectra of NGC 5548. Here epoch A is shown in the left panel, epoch B is shown in the middle panel and epoch C is shown in the right panel. Data from FPMA and FPMB are shown in blue and yellow respectively.}
\label{figure-9}
\end{figure*}
\begin{figure*}
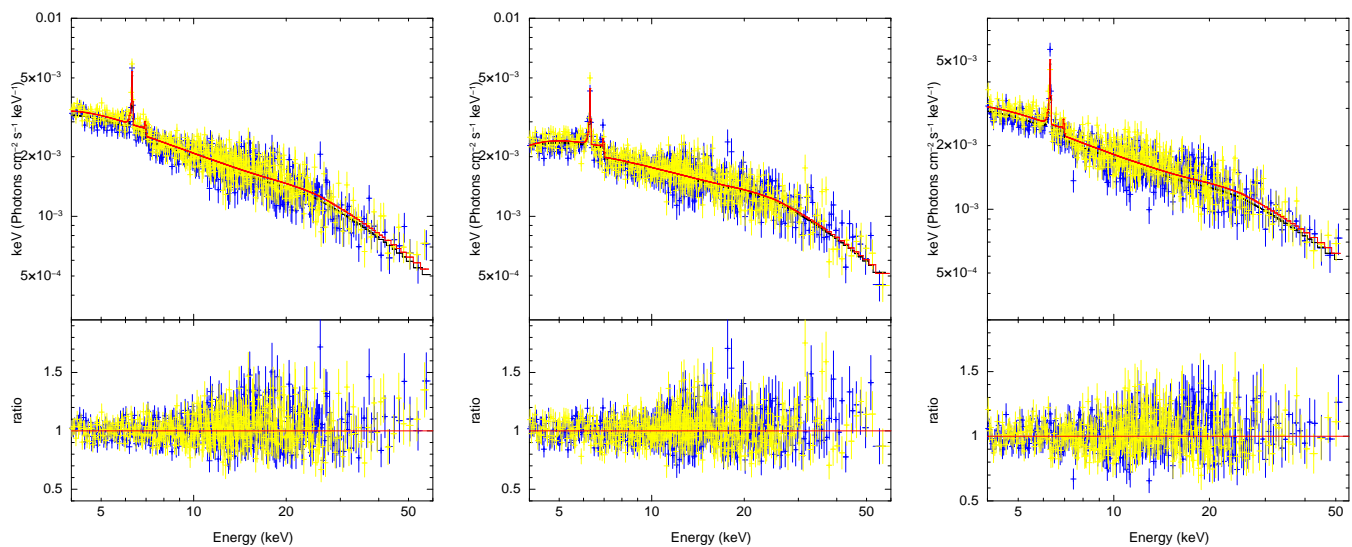

\vbox{
     \hspace{-0.2cm}\includegraphics[scale=0.30]{x06_5548.eps}
     \hspace{-0.2cm}\includegraphics[scale=0.30]{x08_5548.eps}
     \hspace{-0.2cm}\includegraphics[scale=0.30]{x92_5548.eps}
     }
\vspace{-0.5cm}
\caption{The unfolded spectra with the {\it const*TBabs*zTBabs(xillverCP)} model fits and the data to model ratio to the spectra of NGC 5548. Here the left, middle and right panels show the results for epochs D, E and F respectively. Data from FPMA and FPMB are shown in blue and yellow respectively.}
\label{figure-10}
\end{figure*}

\subsection{Physical model fits}
\subsubsection{xillver and xillverCP}
Accretion disk reflection features in the form of narrow FeK$\alpha$ line was 
conspicuously present in the residuals of NGC 3227, NGC 5548 and epoch D spectra of MR 2251$-$178 and the line was modelled using a gaussian component in {\it pexrav} model fits. For MR 2251$-$178 we have seen signatures of reflection (though weak) in few epochs. This is evident from the best fit values of the reflection fraction obtained using {\it pexrav} (see Table \ref{table-3}) and in the residuals of the {\it zpo} model fits (see Fig. \ref{figure-4}: top panel). We therefore modelled the spectra of the sources with the self consistent ionized reflection model {\it xillver}\citep{2010ApJ...718..695G,2013ApJ...768..146G}. The emergent reflected spectrum from the surface of the X-ray illuminated accretion disk is calculated in the model by solving the equations of radiative transfer, energy balance and ionization equilibrium in a Compton thick plane parallel medium \citep{2013ApJ...768..146G}. The model fit to the spectra has
the following form in XSPEC,
\begin{equation}
const*TBabs(xillver)
\end{equation}
Here also, to account for the effect of intrinsic absorption, {\it zTBabs} was used for all epochs of NGC 5548, three epochs (G, H and I) of NGC 3227 and epoch E of MR 2251$-$178. The intrinsic hydrogen column densities,$N_{\rm H}$(zTBabs) were kept as  free parameters. In this model fit, the parameters describing the properties of the corona such as $\Gamma$, $\rm{E_{cut}}$ and R were made to vary, while the inclination angle was frozen to the  value of 50$^{\circ}$ for 
NGC 3227, 30$^{\circ}$ for NGC 5548 and 60$^{\circ}$ for MR 2251$-$178. The other 
parameter that was kept free is the normalization. To account for a fair comparison between the $\rm{E_{cut}}$ values obtained from {\it pexrav} and {\it xillver} the reflector was assumed to be neutral i.e. log$\zeta$ = 0 erg cm s$^{-1}$. Treating the ionization as a free parameter too, returned results consistent with the ones obtained using log$\zeta$ = 0 erg cm s$^{-1}$. The value of Fe abundance was fixed to the solar value. 
The results of the model fits are given in Tables  \ref{table-2},  \ref{table-3} and \ref{table-4}. 
Fitting the spectra using {\it xillver} yielded values of $\rm{E_{cut}}$ similar to that obtained using {\it pexrav}, although the errors in the $\rm{E_{cut}}$ measurements using {\it xillver} are relatively smaller.

As we are interested in the determination of $\rm{kT_{e}}$ and its variation we decided to use the thermal Comptonization model {\it xillverCP} \citep{2014ApJ...782...76G}. This model that takes into account the emission lines assuming it to be originating from disk reflection includes the Comptonization  model {\it nthcomp} \citep{1996MNRAS.283..193Z,1999MNRAS.309..561Z} plus the ionized reflection model {\it xillver}\citep{2010ApJ...718..695G,2013ApJ...768..146G}. The model fit to the spectra has
the following form in XSPEC,
\begin{equation}
const*TBabs(xillverCP)
\end{equation}
We used {\it xillverCP} to model the primary continuum and the reflection spectrum simultaneously. Also, to take care of the intrinsic absoption present in all epochs of NGC 5548, epoch G, H and I of NGC 3227 and epoch E of MR 2251$-$178 spectra the {\it zTBabs} component was added with the described model. The results of the model fits are given in Tables  \ref{table-2},  \ref{table-3} and \ref{table-4}. 

\section{Results and Discussion}
Here, we aimed to find changes of $\rm{kT_{e}}$ in NGC 3227, NGC 5548 and MR 2251$-$178 . We 
discuss below the results obtained on them.

\begin{table*}
\caption{Results of the model fits to the spectra of NGC 5548. The models are Model I: {\it const*TBabs*zTBabs(zpo+zgauss)}, Model II: {\it const*TBabs*zTBabs(pexrav+zgauss)}, Model III: {\it const*TBabs*zTBabs(xillver)} and Model IV: {\it const*TBabs*zTBabs(xillverCP)}. The width of the FeK$\alpha$ line was frozen to the value of 0.1 keV during the fitting. The fluxes are in units of 10$^{-10}$ erg cm$^{-2}$ s$^{-1}$ in the 4$-$60 keV band except in epoch F where the flux was derived in 4$-$55 keV band. Columns and parameters have the same meaning as given in Table \ref{table-2}}.\label{table-4}
\centering
\begin{tabular}{p{0.12\linewidth}p{0.12\linewidth}p{0.10\linewidth}p{0.10\linewidth}p{0.10\linewidth}p{0.10\linewidth}p{0.10\linewidth}}
\hline
Parameter & epoch A & epoch B & epoch C & epoch D & epoch E & epoch F\\
\hline\hline
\multicolumn{7}{c}{Model I : {\it const*TBabs*zTBabs(zpo+zgauss)}} \\
\hline
$\Gamma$ & 1.62$^{+0.03}_{-0.03}$ & 1.60$^{+0.03}_{-0.03}$ & 1.59$^{+0.02}_{-0.02}$ & 1.64$^{+0.02}_{-0.02}$ & 1.57$^{+0.03}_{-0.02}$ & 1.57$^{+0.03}_{-0.01}$  \\ 
$\rm{N_H}$(zTBabs) & 3.89$^{+1.76}_{-1.73}$ & $<3.21$ & 6.29$^{+1.40}_{-1.32}$ & 2.70$^{+1.28}_{-1.27}$ & 8.97$^{+1.65}_{-1.63}$ & $<$1.28 \\
E & 6.37$^{+0.09}_{-0.09}$ & 6.33$^{+0.08}_{-0.08}$ & 6.32$^{+0.04}_{-0.04}$ & 6.38$^{+0.04}_{-0.04}$ & 6.66$^{+0.04}_{-0.04}$ & 6.31$^{+0.04}_{-0.04}$ \\
EW & 70$^{+42}_{-23}$ & 89$^{+27}_{-31}$ & 117$^{+34}_{-34}$ & 107$^{+28}_{-29}$ & 114$^{+26}_{-44}$ & 137$^{+24}_{-45}$ \\
norm & 100$^{+10}_{-9}$ & 86$^{+11}_{-11}$ & 79$^{+8}_{-8}$ & 93$^{+7}_{-6}$ &  77$^{+6}_{-5}$ & 67$^{+5}_{-2}$\\
$\chi^2/dof$ & 684/643 & 612/634 & 880/811 & 890/813 & 840/810 & 692/692 \\
$\rm{C_{FPMA/FPMB}}$ & 1.02$^{+0.02}_{-0.02}$ & 1.05$^{+0.01}_{-0.01}$ & 1.02$^{+0.01}_{-0.01}$ & 1.05$^{+0.01}_{-0.01}$ & 1.02$^{+0.01}_{-0.01}$ & 1.05$^{+0.01}_{-0.01}$\\
\hline
\multicolumn{7}{c}{Model II : {\it const*TBabs*zTBabs(pexrav+zgauss)}} \\
\hline
$\Gamma$ & 1.75$^{+0.02}_{-0.10}$ & 1.81$^{+0.02}_{-0.08}$ & 1.62$^{+0.10}_{-0.10}$ & 1.79$^{+0.09}_{-0.09}$  & 1.61$^{+0.10}_{-0.11}$ & 1.81$^{+0.08}_{-0.09}$\\
$\rm{N_H}$(zTBabs) & 6.06$^{+1.58}_{-2.46}$ & 4.77$^{+1.31}_{-2.05}$ & 5.61$^{+2.34}_{-2.36}$ & 4.46$^{+2.13}_{-2.16}$ & 6.91$^{+2.45}_{-2.47}$ & 3.66$^{+1.18}_{-2.10}$ \\
$\rm{E_{cut}}$ & $>$345 & $>$502 & 152$^{+158}_{-54}$ & $>$170 & 160$^{+178}_{-58}$ & $>$414 \\
R & 0.19$^{+0.14}_{-0.11}$  & 0.35$^{+0.16}_{-0.14}$ & 0.28$^{+0.11}_{-0.10}$ & 0.37$^{+0.12}_{-0.11}$ & 0.33$^{+0.12}_{-0.10}$ & 0.42$^{+0.14}_{-0.13}$\\
E & 6.37$^{+0.10}_{-0.10}$ & 6.32$^{+0.09}_{-0.10}$ & 6.32$^{+0.04}_{-0.04}$ & 6.38$^{+0.05}_{-0.05}$ & 6.35$^{+0.04}_{-0.04}$ & 6.31$^{+0.05}_{-0.05}$ \\
EW & 55$^{38}_{-27}$ & 64$^{+31}_{-46}$ & 114$^{+24}_{-38}$ & 89$^{+25}_{-20}$ & 113$^{+29}_{-34}$ & 104$^{+44}_{-27}$ \\
norm & 126$^{+7}_{-23}$ & 125$^{+15}_{-19}$ & 79$^{+18}_{-15}$ & 118$^{+23}_{-20}$ &  73$^{+17}_{-14}$ & 103$^{+3}_{-17}$ \\
$\chi^2/dof$ & 675/641 & 588/632 & 834/809 & 840/811 & 780/808 & 652/690 \\
$\rm{C_{FPMA/FPMB}}$ & 1.02$^{+0.02}_{-0.02}$ & 1.05$^{+0.01}_{-0.01}$ & 1.02$^{+0.01}_{-0.01}$ & 1.05$^{+0.01}_{-0.01}$ & 1.02$^{+0.01}_{-0.01}$ & 1.05$^{+0.01}_{-0.01}$ \\
\hline
\multicolumn{7}{c}{Model III : {\it const*TBabs*zTBabs(xillver)}} \\
\hline
$\Gamma$ & 1.75$^{+0.02}_{-0.02}$ & 1.78$^{+0.02}_{-0.04}$ & 1.68$^{+0.02}_{-0.01}$ & 1.77$^{+0.01}_{-0.01}$ & 1.66$^{+0.01}_{-0.02}$ & 1.81$^{+0.01}_{-0.01}$ \\
$\rm{N_H}$(zTBabs) & 6.47$^{+0.98}_{-0.91}$ & 4.83$^{+0.95}_{-0.91}$ & 7.13$^{+0.73}_{-0.73}$ & 4.52$^{+0.69}_{-0.67}$ & 8.16$^{+0.76}_{-0.75}$ & 4.25$^{+0.83}_{-0.81}$ \\
$\rm{E_{cut}}$ & $>$487 & $>$480 & 129$^{+15}_{-13}$ & 179$^{+40}_{-23}$ & 133$^{+15}_{-13}$ & $>$395 \\
R & 0.25$^{+0.08}_{-0.08}$ & 0.36$^{+0.10}_{-0.09}$ & 0.49$^{+0.09}_{-0.08}$ & 0.49$^{+0.08}_{-0.08}$ & 0.51$^{+0.09}_{-0.08}$ & 0.54$^{+0.11}_{-0.10}$ \\
norm & 4.35$^{+0.10}_{-0.06}$ & 3.61$^{+0.47}_{-0.14}$ & 2.20$^{+0.03}_{-0.03}$ & 2.47$^{+0.03}_{-0.03}$ & 2.07$^{+0.04}_{-0.04}$ & 2.70$^{+0.04}_{-0.04}$ \\
$\chi^2/dof$ & 677/643 & 595/634 & 851/811 & 849/813 & 792/810 & 666/692 \\
$\rm{C_{FPMA/FPMB}}$ & 1.02$^{+0.02}_{-0.02}$ & 1.05$^{+0.01}_{-0.01}$ & 1.02$^{+0.01}_{-0.01}$ & 1.05$^{+0.01}_{-0.01}$ & 1.02$^{+0.01}_{-0.01}$ & 1.05$^{+0.01}_{-0.01}$ \\
\hline
\multicolumn{7}{c}{Model IV : {\it const*TBabs*zTBabs(xillverCP)}} \\
\hline
$\Gamma$ & 1.77$^{+0.02}_{-0.02}$ & 1.80$^{+0.02}_{-0.03}$ & 1.79$^{+0.01}_{-0.01}$ & 1.85$^{+0.01}_{-0.01}$ & 1.77$^{+0.01}_{-0.01}$ & 1.84$^{+0.01}_{-0.02}$ \\
$\rm{N_H}$(zTBabs) & 6.47$^{+0.99}_{-0.91}$ & 4.63$^{+0.94}_{-0.92}$ & 8.72$^{+0.73}_{-0.74}$ & 5.66$^{+0.69}_{-0.68}$ & 9.79$^{+0.76}_{-0.75}$ & 4.30$^{+0.87}_{-0.92}$ \\
$\rm{kT_{e}}$ & $>$53 & $>$54 & 39$^{+14}_{-10}$ & 65$^{+147}_{-24}$ & 38$^{+12}_{-9}$ & $>$65  \\
R & 0.25$^{+0.08}_{-0.09}$ & 0.35$^{+0.11}_{-0.08}$ & 0.43$^{+0.08}_{-0.08}$ & 0.46$^{+0.09}_{-0.08}$ & 0.45$^{+0.08}_{-0.08}$ & 0.55$^{+0.11}_{-0.11}$  \\
norm & 3.46$^{+0.05}_{-0.05}$ & 2.95$^{+0.34}_{-0.04}$ & 2.23$^{+0.03}_{-0.03}$ & 2.46$^{+0.03}_{-0.04}$ & 2.07$^{+0.03}_{-0.03}$ & 2.34$^{+0.08}_{-0.03}$ \\
$\chi^2/dof$ & 677/643 & 595/634 & 855/811 & 852/813 & 795/810 & 667/692  \\
$\rm{C_{FPMA/FPMB}}$ & 1.02$^{+0.02}_{-0.02}$ & 1.05$^{+0.01}_{-0.01}$ & 1.02$^{+0.01}_{-0.01}$ & 1.05$^{+0.01}_{-0.01}$ & 1.02$^{+0.01}_{-0.01}$ & 1.05$^{+0.01}_{-0.01}$ \\
Flux & 1.23$^{+0.01}_{-0.01}$ & 1.11$^{+0.01}_{-0.01}$ & 1.02$^{+0.01}_{-0.01}$ & 1.05$^{+0.01}_{-0.01}$ & 0.96$^{+0.01}_{-0.01}$ & 0.92$^{+0.01}_{-0.01}$  \\
\hline\hline
\end{tabular}
\end{table*}

\begin{figure*}
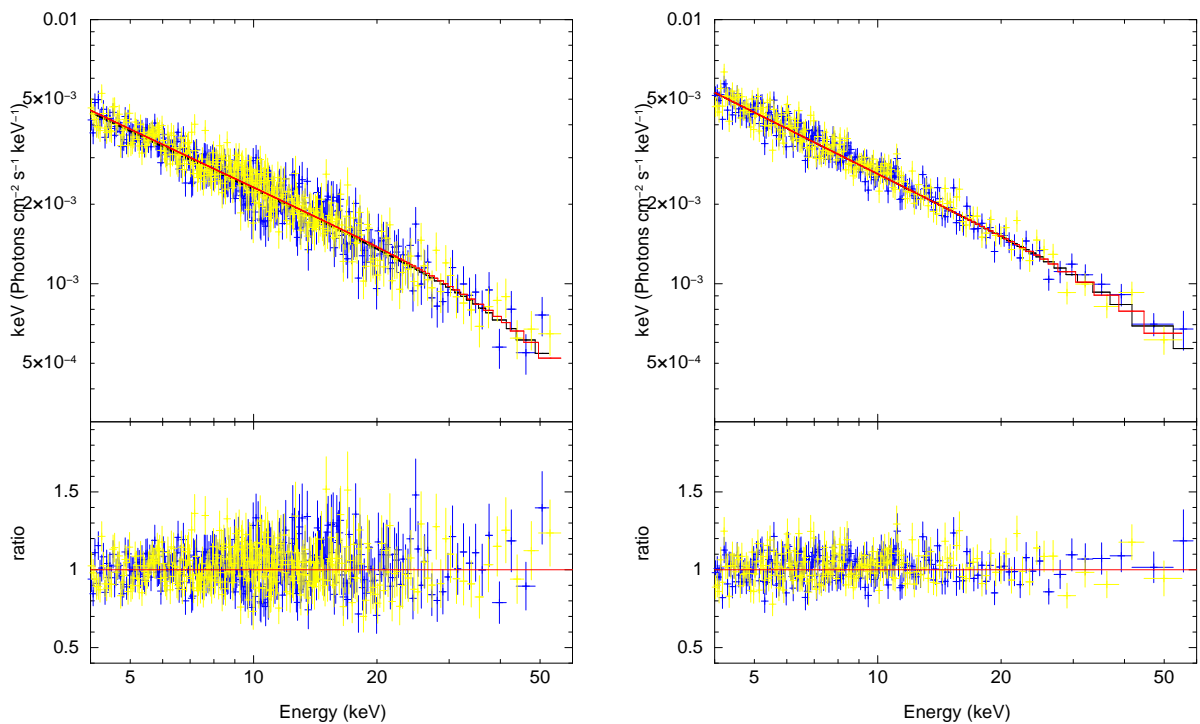

\vbox{
     \hspace{-0.05cm}\includegraphics[scale=0.40]{x02_mr.eps}
     \hspace{-0.05cm}\includegraphics[scale=0.40]{x04_mr.eps}
     }
     \vspace{-0.5cm}
     \caption{The unfolded spectra along with the data to model ratio for {\it const*TBabs(xillverCp)} fit to epoch A (left panel) and epoch B (right panel) observations of MR 2251$-$178. Here blue points are for FPMA and yellow points are for FPMB.}
\label{figure-11}
\end{figure*}
\vspace{-0.5cm}
\begin{figure*}
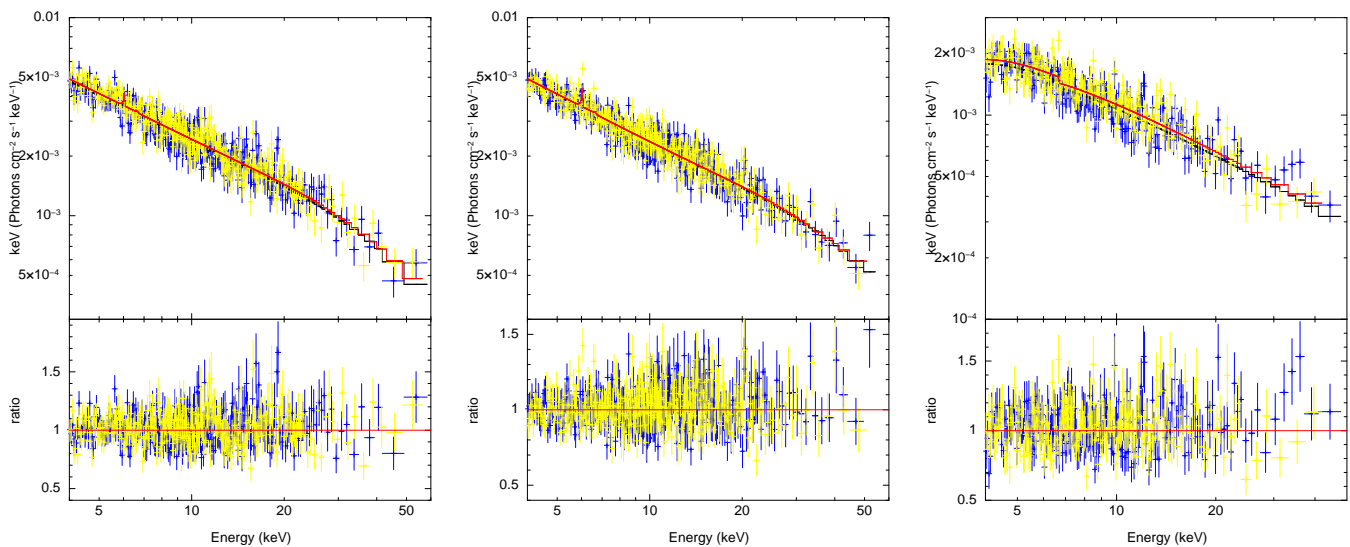

\vbox{
     \hspace{-0.2cm}\includegraphics[scale=0.30]{x06_mr.eps}
     \hspace{-0.2cm}\includegraphics[scale=0.30]{x08_mr.eps}
     \hspace{-0.2cm}\includegraphics[scale=0.30]{x92_mr.eps}
     }
\vspace{-0.5cm}
\caption{The unfolded spectra along with the data to model ratio for {\it const*TBabs(xillverCp)} fit to epoch C (left panel), epoch D (middle panel) and epoch E (right panel) observations of MR 2251$-$178. Here blue points are for FPMA and yellow points are for FPMB. In epoch E the {\it zTBabs} component was added to the model.}
\label{figure-12}
\end{figure*}

\subsection{NGC 3227}
{\it NuSTAR} observed this source nine times between November 2016 and 
December 2019. Of these, results on seven epochs were reported by \cite{2021MNRAS.502...80K}, wherein they were
able to constrain $\rm{E_{cut}}$ in three epochs using phenomenological fits and $\rm{kT_{e}}$ 
in two epochs using physical model fits. Here, we report results for two additional 
epochs and for all the nine epochs, we used  both phenomenological
and physical model fits to model the spectra. 

Ratios of the model {\it const*TBabs(zpo + zgauss)}, {\it const*TBabs(pexrav + zgauss)} and {\it const*TBabs(xillverCP)} fits to the observed FPMA and FPMB spectra carried out on the OBSID 60202002002, the one with the longest exposure time are shown in Fig. \ref{figure-2}. Weak FeK$\alpha$ line is present in all the OBSIDs and therefore in the {\it const*TBabs(zpo)} and {\it const*TBabs(pexrav)} model fits we included a {\it zgauss} component to account for the line.  
All the above models could fit the spectra of all epochs of NGC 3227
reasonably well except for epochs G, H and I, where the $\chi^2/dof$ was 
greater than 1.2 and there is evidence for the presence of 
significant absorption at the low energy end (see Fig. \ref{figure-5}). Addition of an extra absorption component 
{\it zTBabs} that takes into account the effects of host
galaxy absorption to all the models improved the fit significantly
(see left, middle and right panel of Fig. \ref{figure-8}).  The {\it const*TBabs(pexrav + zgauss)} 
model fit to the FPMA/FPMB spectra showed evidence for variation in $\rm{E_{cut}}$. To analyze the variation of $\rm{E_{cut}}$ with time as well as to find the relation between $\rm{E_{cut}}$ and $\rm{kT_{e}}$ we used the $\rm{E_{cut}}$ values obtained from {\it xillver} model fits. The lowest value of $\rm{E_{cut}}$ of 92$^{+10}_{-8}$ keV was obtained in epoch B and the 
highest value of $\rm{E_{cut}}$ was observed in
epoch E (378$^{+152}_{-91}$ keV) while intermediate values of 
$\rm{E_{cut}}$ were obtained during other epochs. These observed variation of $\rm{E_{cut}}$ indicates that the coronal temperature of NGC 3227 must be changing with time. This 
is also evident in Fig. \ref{figure-6} (upper panel) where 
the values of $\rm{E_{cut}}$ are plotted against epochs. To quantify the significance of the variation 
of $\rm{E_{cut}}$ we fitted a constant (mean of all the $\rm{E_{cut}}$ measurements)
to the variation of $\rm{E_{cut}}$ with epoch and calculated the $\chi^2$ and the 
null hypothesis probability (p) that $\rm{E_{cut}}$ does not change with time (shown 
as a dashed line in the top panel of Fig. \ref{figure-6}). We found $\chi^2$/dof $>>$ 10, and a 
$p$ value of 0.0 that the $\rm{E_{cut}}$ does not change with time. The values of $\rm{E_{cut}}$ 
obtained during the first two epochs (A and B) are similar to the value of 
$\rm{E_{cut}}$ = 90$\pm$20 keV reported by  \cite{2009ApJ...691..922M}. 
\cite{2013ApJ...763..111V} from an analysis of XMM and BAT spectra pointed 
$\rm{E_{cut}}$ to lie outside the BAT range at $>$ 636 keV. Recently \cite{2021MNRAS.502...80K} too reported $\rm{E_{cut}}$ values for epoch A, B and G, and lower limits for epochs C,D,E and F from {\it pexrav} model fits. The values of $\rm{E_{cut}}$ obtained here using {\it xillver} for epochs A, B and G are in agreement with that of \cite{2021MNRAS.502...80K} from using {\it pexrav} model. Also, our model fits could constrain $\rm{E_{cut}}$ during epochs E and F using {\it xillver}.

Comptonization model fits using {\it xillverCP} provided values of $\rm{kT_{e}}$ which too was found to vary between epochs. We could constrain $\rm{kT_{e}}$ for epochs A, B, C and G, and obtain lower limits for epochs D, E, F, H and I. \cite{2021MNRAS.502...80K} using the same model used here could constrain $\rm{kT_{e}}$ for only epochs A and B. Our results for epochs A and B are in agreement with that of \cite{2021MNRAS.502...80K}. The variation of $\rm{kT_{e}}$ is shown in the lower panel of Fig. \ref{figure-6}. The trend seen in the variation of $\rm{kT_{e}}$ with epochs is 
similar to the variation of $\rm{E_{cut}}$. From  Comptonization model fits we found 
the lowest value of $\rm{kT_{e}}$ = 
28$^{+9}_{-6}$ keV during epoch B while the highest value of 
$\rm{kT_{e}}$ = 56$^{+131}_{-18}$ keV was obtained for epoch E. We obtained lower limits of $\rm{kT_{e}}$ of 80, 45, 47, 85 and 36 keV during epochs D, E, F, H and I respectively. Spectral fits using {\it xillverCP} along with the data to model ratio for all the epochs are shown in Figures \ref{figure-7} and \ref{figure-8}. The results of the model fits are given in Table \ref{table-2}. To quantify the significance of the variation of $\rm{kT_{e}}$ we fitted a constant (mean of all the $\rm{kT_{e}}$ measurements) to the variation of $\rm{kT_{e}}$ with epoch and calculated the $\chi^2$ and the 
null hypothesis probability (p) that $\rm{kT_{e}}$ does not change with time (shown as a dashed line in Fig. \ref{figure-6}). We found $\chi^2$/dof of 12.86, and a $p$ value of 0.005 that $\rm{kT_{e}}$ does not change with time. The contour plots between $\Gamma$ against $\rm{kT_{e}}$ and R against $\rm{kT_{e}}$ for NGC 3227 for all the epochs are shown in Fig. \ref{figure-14} and \ref{figure-15} (left panels).From the contour plots it is evident that the coronal temperature variation is prominent in NGC 3227. We therefore, conclude that we found variation in the temperature of the corona in NGC 3227. According to \cite{2001ApJ...556..716P} for an optically thick corona ($\tau$ $>$ 1) $\rm{E_{cut}}$ = 3$\rm{kT_{e}}$. However, the relation between $\rm{E_{cut}}$ and $\rm{kT_{e}}$ can be complicated in the case of a non-static corona such as the one with outflows \citep{2014ApJ...783..106L}. Also, according to \cite{2019A&A...630A.131M}, the relation of $\rm{E_{cut}}$ = 2$-$3 $\rm{kT_{e}}$ is valid only for low values of $\tau$ and $\rm{kT_{e}}$. For NGC 3227 using all the five epochs in which we could constrain both $\rm{E_{cut}}$ and $\rm{kT_{e}}$ we found $\rm{E_{cut}}$ = 3.94$\pm$0.62 $\rm{kT_{e}}$ which is similar to that of \cite{2001ApJ...556..716P} and \cite{2019A&A...630A.131M}

\begin{table*}
\caption{Results of the correlation analysis between different parameters for NGC 3227. Provided are the slope (m), intercept (c), Pearson's correlation coefficient (r) and the probability (p) for null hypothesis (no correlation) from OLS fit and the least squares fit from simulated points. See text in Section 4.4.}\label{table-5}
\centering
\begin{tabular}{lrrrcrrrc}
\hline
Parameter & \multicolumn{4}{c}{OLS} & \multicolumn{4}{c}{Simulated} \\ \\
\hline
& m & c & r & p & m & c & r & p  \\
\hline
$\Gamma$/Flux & -0.07$\pm$0.03 & 1.90$\pm$0.03 & -0.63 & 0.07 & -0.07$\pm$0.03 & 1.91$\pm$0.04 & -0.64 & 0.06 \\
$\rm{kT_{e}}$/Flux & 18$\pm$28 & 22$\pm$31 & 0.42 & 0.58 & 18$\pm$150 & 13$\pm$169 & 0.08 & 0.52 \\
$\Gamma$/$\rm{kT_{e}}$ & 0.001$\pm$0.0006 & 1.75$\pm$0.03 & 0.86 & 0.14 & 0.0005$\pm$0.0002 & 1.78$\pm$0.02 & 0.80 & 0.20 \\
$R$/$\rm{kT_{e}}$ & -0.005$\pm$0.005 & 1.03$\pm$0.22 & -0.57 & 0.43 & -0.001$\pm$0.002 & 0.90$\pm$0.17 & -0.36 & 0.60 \\
$R$/Flux & -0.51$\pm$0.13 & 1.31$\pm$0.13 & -0.84 & 0.01 & -0.04$\pm$0.15 & 0.78$\pm$0.16 & -0.09 & 0.68 \\
$\tau$/Flux & -0.29$\pm$0.62 & 2.35$\pm$0.65 & -0.17 & 0.66 & -0.98$\pm$1.88 & 3.15$\pm$2.12 & -0.35 & 0.65 \\
$y$/Flux & -0.15$\pm$0.70 & 13.06$\pm$0.79 & -0.15 & 0.85 & -0.53$\pm$1.50 & 14.01$\pm$1.71 & -0.21 & 0.66 \\
$\tau$/$\rm{kT_{e}}$ & -0.043$\pm$0.002 & 4.19$\pm$0.09 & -0.99 & 0.00 & -0.02$\pm$0.00 & 3.44$\pm$0.28 & -0.97 & 0.03 \\
\hline\hline
\end{tabular}
\end{table*}

\subsection{MR 2251$-$178}
\begin{figure*}
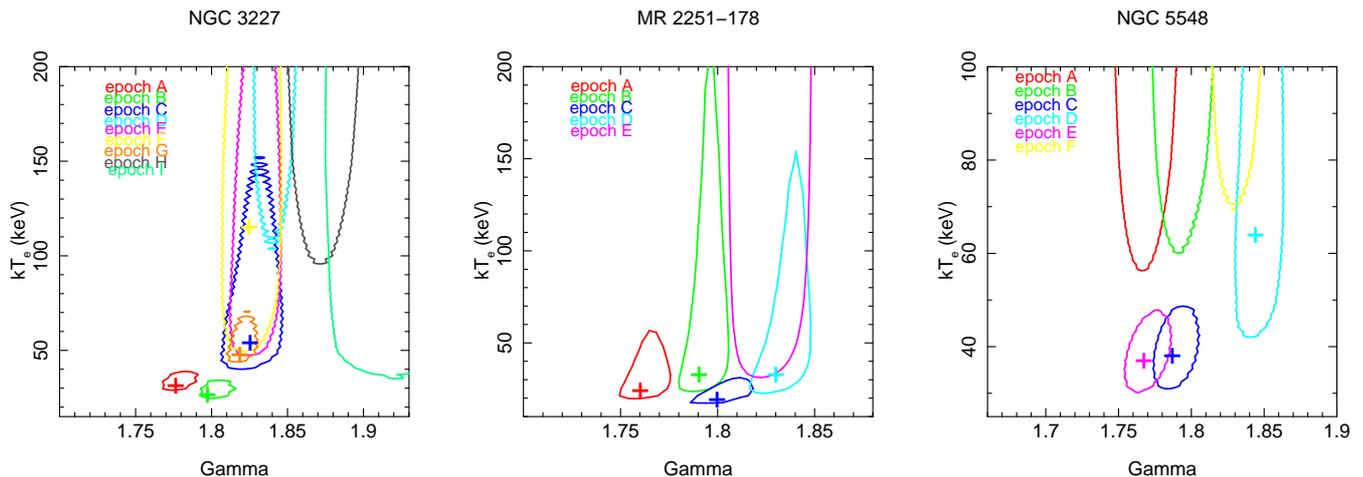

\vbox{
     \hspace{-0.4cm}\includegraphics[scale=0.58]{xg_ngc3227_new.eps}
     \hspace{-0.4cm}\includegraphics[scale=0.58]{xg_mr_new.eps}
     \hspace{-0.4cm}\includegraphics[scale=0.58]{xg_ngc5548_new.eps}
    }
\vspace{-0.5cm}
\caption{The 90 per cent confidence level contours between $\rm{kT_{e}}$ and 
	$\Gamma$ for the {\it xillverCP} model fit to the FPMA/FPMB spectra of 
	NGC 3227(left), MR 2251$-$178(middle) and NGC 5548(right). The colours indicate different OBSIDs. Here for MR 2251$-$178 the limits of $\rm{kT_{e}}$ do not have a correspondence with the values given in Table \ref{table-3}. This is because the contours were generated by freezing the $R$ parameter so as to put all the five OBSIDs together.}
\label{figure-14}
\end{figure*}

\begin{figure*}
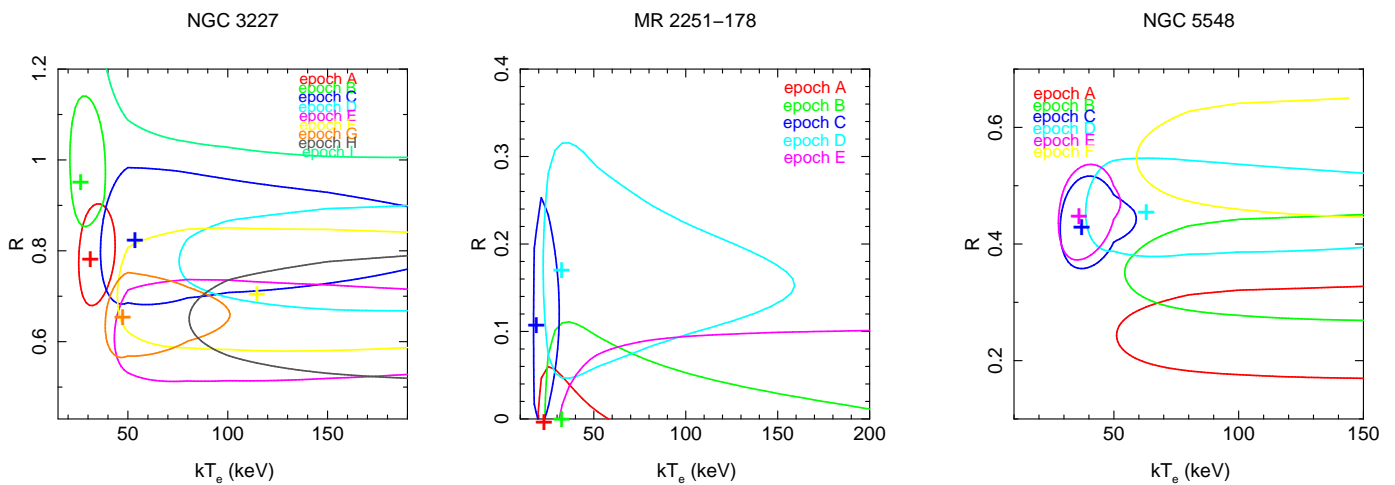

\vbox{
     \hspace{-0.4cm}\includegraphics[scale=0.58]{xr_ngc3227_new.eps}
     \hspace{-0.4cm}\includegraphics[scale=0.58]{xr_mr_new.eps}
     \hspace{-0.4cm}\includegraphics[scale=0.58]{xr_ngc5548_new.eps}
    }
\vspace{-0.5cm}
\caption{The 90 per cent confidence level contours between R and $\rm{kT_{e}}$ for the {\it xillverCP} model fit to the FPMA/FPMB spectra of NGC 3227(left), MR 2251$-$178(middle) and NGC 5548(right).}
\label{figure-15}
\end{figure*}

This source has five epochs of observations that are public and having exposure $>$ 20 ks. In this work we analyzed all of them. Simple power law fits to the FPMA spectra of all the epochs is shown in the right panel of Fig. \ref{figure-1}. From this figure, noticeable 
change in the spectra could not be ascertained. The ratio of the model fits  {\it const*TBabs(zpo)}, {\it const*TBabs(pexrav)} and {\it const*TBabs(xillverCP)} to the observed FPMA and FPMB spectra on OBSID 90601637002, the one with the longest exposure time are shown in Fig. \ref{figure-4}. From two sets of observations from {\it Einstein} separated by about a year \cite{1984ApJ...281...90H} found evidence of variable X-ray absorption 
in  MR 2251$-$178, with the column density changing from  
$<$ 5 $\times$ 10$^{21}$ cm$^{-2}$ to 2 $\times$ 10$^{22}$ cm$^{-2}$, 
suggested to be due to the presence of a warm absorber. {\it EXOSAT} and 
{\it Ginga} observations revealed strong correspondence between the absorbing 
column density and the flux of the source with the low energy absorption 
decreasing with the increasing flux of the  source. These observations were 
explained by variable absorption in photo-ionized gas along the line of sight 
\citep{1993MNRAS.262..817M,1990MNRAS.242..177P}.  However, 
\cite{1992A&A...266...57W} from an analysis of the {\it EXOSAT} data argued 
that the variability seen in the source can be explained without invoking the 
presence of a warm absorber. From the ratio of the observed data to the model fit we did not find any signature of absorption that could affect the source spectra in all observations but in epoch E and to take care of this we added the {\it zTBabs} component to all the four models in epoch E(see Fig \ref{figure-11} and \ref{figure-12}). Model fits to the four sets of observations that span about five years using {\it const*TBabs(zpo)} do not reveal the presence of FeK$\alpha$ line in the  spectra, the reflection bump was also found to be either negligible or weak (see top panel of Fig. \ref{figure-4}), more likely due to poor S/N. However, the ratio plot for the model {\it const*TBabs(zpo)} to the epoch D spectra revealed the presence of the FeK$\alpha$ line at around 6.4 keV. We therefore added a {\it zgauss} component with the model and found the energy of the line at 6.49$^{+0.37}_{-0.27}$ keV with a fixed width of 0.1 keV, letting the parameter free did not significantly improve the fit. For this source previously the FeK$\alpha$ line was reported to be present in the {\it Ginga} observations with an equivalent width of 125$^{+100}_{-105}$ eV \citep{1993MNRAS.262..817M}. Relatively strong FeK$\alpha$ was also reported to be present in the {\it BeppoSaX} observation \citep{2001A&A...376..413O} and a narrow FeK$\alpha$ line was present in the {\it Chandra} observations \citep{2005ApJ...627...83G}. From {\it BeppoSAX} observations in the 0.1$-$200 keV band, \cite{2001A&A...376..413O} found a $\rm{E_{cut}}$ value of around 100 keV which is similar to that obtained here.
\begin{figure}
\vbox{
      \includegraphics[scale=0.48]{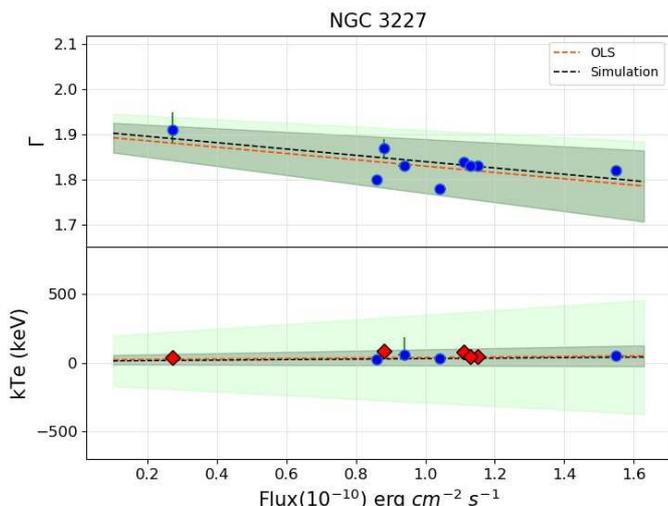}
      }
\vspace{-0.34cm}
\caption{Correlation of $\Gamma$ and $\rm{kT_{e}}$ with the X-ray brightness of 
	NGC 3227. The orange dashed line is the ordinary linear
	least squares (OLS) fit to the data points. The black dashed line is the linear line drawn using the median values of the simulated parameters. The grey shaded region indicates the upper and lower errors in the fit parameters for OLS and the green region indicates the errors in the fit parameter obtained from the simulation. In the least square analysis, epochs in which we were unable to constrain $\rm{kT_{e}}$ were dropped (indicated with red point).} \label{figure-13}
\end{figure}
To find evidence for the change in $\rm{kT_{e}}$ if any, we carried out 
fitting of the observations with the physical model, {\it const*TBabs(xillverCP)}.  
Examination of the results of the fit (Table \ref{table-3}) shows that
the $\rm{kT_{e}}$ obtained during all the epochs agree 
within errors. Though we could not find any signature of $\rm{kT_{e}}$ variation with epochs from the {\it xillverCP} fit, the $\rm{E_{cut}}$ values obtained from the {\it xillver} model fit during epochs A, C and D agree within errors except that of epoch B. This could be due to the
quality of the data in epoch B, as the values of $\rm{E_{cut}}$ and $\rm{kT_{e}}$ obtained during that epoch also have large error bars. To test for the $\rm{kT_{e}}$ variation if any, we plotted the 90 per cent contours between $\rm{kT_{e}}$ and $\Gamma$. The 90 percent contours of $\rm{kT_{e}}$ against $\Gamma$ overlap (see Fig. \ref{figure-14}) and we conclude that in MR 2251$-$178 we did not find any variation of $\rm{kT_{e}}$ with time. The 90 per cent contours of R against $\rm{kT_{e}}$ are also shown in the bottom panel of Fig. \ref{figure-15}. Considering all the four epochs where $\rm{E_{cut}}$ and $\rm{kT_{e}}$ could be constrained we found $\rm{E_{cut}}$ = 4.84$\pm$0.11 $\rm{kT_{e}}$. This is deviant from the generally adopted $\rm{E_{cut}}$ = 2$-$3 $\rm{kT_{e}}$ \citep{2001ApJ...556..716P}. It is likely the relation between $\rm{E_{cut}}$ and $\rm{kT_{e}}$ is complex and may depend on other physical properties of the sources. Homogeneous analysis of a large number of sources are needed to establish the relation between $\rm{E_{cut}}$ and $\rm{kT_{e}}$ as well as on its dependence on other physical properties. Spectral fits using {\it xillverCP} along with the residuals for all the epochs are shown in Fig. \ref{figure-11} and \ref{figure-12}.


\begin{figure}
     \includegraphics[scale=0.45]{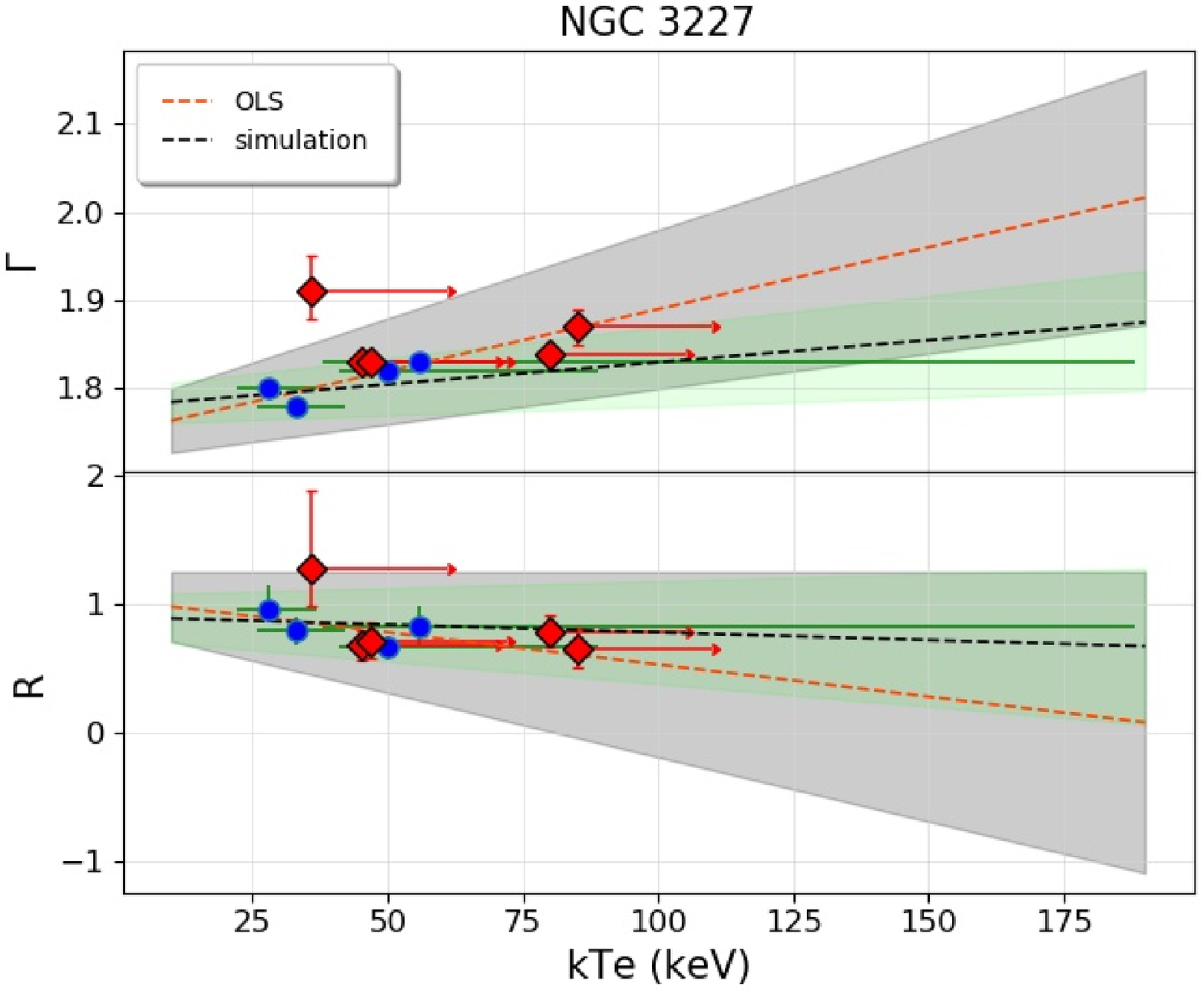}
\caption{Correlation between $\Gamma$ and $R$ with $\rm{kT_{e}}$ of 
	NGC 3227. The dashed lines and shaded regions have the same meaning as in Fig. \ref{figure-13}. Epochs in which $\rm{kT_{e}}$ could not be constrained (shown as red points) were not included in the linear least square analysis.} \label{figure-16}
\end{figure}

\subsection{NGC 5548}
NGC 5548 was observed by {\it NuSTAR} six times between July 2013 to January 2021. Of these, \cite{2018ApJ...863...71Z} have reported results for five epochs. In this work we carried out both phenomenological and physical model fits for all the six epochs.  We fit the spectra using {\it const*TBabs(zpo+zgauss)} and  {\it const*TBabs(pexrav+zgauss)} to model the primary power law emission and 
the reflection component along with the FeK$\alpha$ line with the width of the line being frozen to the value of 0.1 keV. From the ratio of the observed data to the model fit using 
{\it TBabs} we found the presence of absorption component in the 
spectra. So unlike the other sources, for NGC 5548, we added an extra 
component {\it zTBabs} in each model to account for the absorption by the host 
galaxy. Moreover, as {\it xillver} self consistently models the reflected spectrum and its associated FeK$\alpha$ line, we also fit the spectra using {\it xillver} and $\rm{E_{cut}}$ values obtained using {\it xillver} model were used to find variation in $\rm{E_{cut}}$. The $\rm{E_{cut}}$ obtained from {\it xillver} model fits are  $>487$ keV, $>480$ keV, 129$^{+15}_{-13}$ keV, 179$^{+40}_{-23}$ keV, 133$^{+15}_{-13}$ and $>$394 keV for epochs A, B, C, D, E and F respectively. The residuals of the fit to the spectra for the epoch D that has the maximum exposure time, for various models are given in Fig. \ref{figure-3} and the results of the fit are given in Table \ref{table-4}. From {\it xillver} model fits, we could constrain $\rm{E_{cut}}$ only for epochs C, D and E, while \cite{2018ApJ...863...71Z} could constrain $\rm{E_{cut}}$ in the four epochs (A, C, D and E). For the epochs in common, though our values of $\rm{E_{cut}}$ are in agreement with \cite{2018ApJ...863...71Z} the results do not match for epoch E. Though \cite{2018ApJ...863...71Z} claims to have detected $\rm{E_{cut}}$ variation in NGC 5548, our analysis could not confirm changes in $\rm{E_{cut}}$. This could be due to differences in the choice of binning and the energy ranges used in both the work. \cite{2015A&A...577A..38U} via the joint fitting of {\it XMM-Newton}, {\it Chandra}, {\it NuSTAR}, and {\it INTEGRAL} data obtained lower limits for $\rm{E_{cut}}$ in all the epochs except for epoch D. Using the simultaneous {\it XMM-Newton} and {\it NuSTAR} data from 2013 campaign \cite{2016A&A...592A..27C} also fitted the 4$-$79 keV epoch A/B , epoch D and epoch E spectra using cut-off power-law and {\it pexmon}. Their values of $\rm{E_{cut}}$ are in agreement with the $\rm{E_{cut}}$ values obtained in this work.

Fitting the physical Comptonization model {\it xillverCP} to ascertain the change in coronal temperature, we got the highest $\rm{kT_{e}}$ value of 65$^{+147}_{-24}$ keV during epoch D, for epochs A, B and F we obtained lower limits of 53, 54 and 71 keV, while for the remaining two epochs C and E, we obtained similar values of 39$^{+14}_{-10}$ keV and 38$^{+12}_{-9}$ keV respectively. The values of $\rm{kT_{e}}$ between epochs agree to each other within error bars. The model fits to the data along with the data to model ratio for all the epochs of observations are shown in Fig. \ref{figure-9} and \ref{figure-10}. The $\rm{E_{cut}}$ values obtained here points to variability. However, the derived $\rm{kT_{e}}$ values are consistent with each other within error bars. The contour plots between $\rm{kT_{e}}$ against $\Gamma$ as well as R against $\rm{kT_{e}}$ are shown in the right most panels of Fig. \ref{figure-14} and \ref{figure-15}. From these plots, the change in $\rm{kT_{e}}$ is not evident. We thus conclude that we have not found evidence for variation in $\rm{kT_{e}}$ in NGC 5548. Considering all the epochs in which we could constrain both $\rm{E_{cut}}$ and $\rm{kT_{e}}$, we found $\rm{E_{cut}}$ = 3.19$\pm$0.32 $\rm{kT_{e}}$ which is in agreement with the empirical relation of \cite{2001ApJ...556..716P}.

\subsection{Correlation between different parameters}
We discuss below the correlation between various parameters. This is restricted to the source NGC 3227 as the temperature of its corona is found to vary in this work. Since the errors in the measured $\rm{E_{cut}}$ and $\rm{kT_{e}}$ values are not symmetric and there is no conventional way to take care of such errors during correlation study we adopted two procedures to find the correlation between various parameters.

In first case, we neglected the corresponding errors and considered only the best fit values of the parameters and performed the ordinary linear least square (OLS) fit between them. The Pearson's correlation coefficient (r) and the null hypothesis probability (p) for no correlation were also estimated using the best fit values.

In the second case, to take into account the non-symmetric errors we simulated $10^{5}$ points from each rectangular box around the best fit values (x,y) with x and y boundaries of ($\rm{x_{low}}$,$\rm{x_{high}}$) and ($\rm{y_{low}}$,$\rm{y_{high}}$) respectively. Here $\rm{x_{low}}$, $\rm{y_{low}}$ and $\rm{x_{high}}$, $\rm{y_{high}}$ refer to the lower and upper errors in x and y values. Linear least squares fit was done for $10^{5}$ times yielding distribution of the slope (m), the intercept (c), the Pearson's rank correlation coefficient and the probability of no correlation. The median values of the distributions were taken to represent the best fit values of the correlation. All the values and the errors for the unweighted as well as for the simulated correlation are given in Table \ref{table-5}.

\subsubsection{$\Gamma$ v/s Flux}
In Seyfert galaxies, the X-ray spectra are generally found to be softer with increasing X-ray flux \citep{2003ApJ...593...96M}. We show in Fig. \ref{figure-13}, the correlation between $\Gamma$ and the brightness of NGC 3227. For the source, each point in the Figure corresponds to $\Gamma$ and flux obtained by {\it xillverCP} model fits to each epoch of spectra. Ordinary linear least squares fit to the data are shown by orange dashed lines. The black dashed lines show the linear least square fit using the median values of the simulated points using the lower and upper errors in $\Gamma$ and flux values. The grey and green shaded regions are the area bounded by the errors in the least square fit parameters. For  NGC 3227, we found anti correlation between $\Gamma$ and flux (see Fig. \ref{figure-13} and Table \ref{table-5}) between epochs of observations significant at the greater than 90 per cent level. 

\subsubsection{$\rm{kT_{e}}$ v/s Flux}
Correlation between the changes in the temperature of the corona with other physical parameters of the sources such as its apparent brightness as well as its spectral shape can provide important
constraints in enhancing our understanding of AGN corona. The correlation between $\rm{kT_{e}}$ and flux is shown in Fig. \ref{figure-13}. Also, shown in the figure are the ordinary and simulated linear least squares fit to the data. No correlation is found between $\rm{kT_{e}}$ and flux in NGC 3227. 

\subsubsection{$\Gamma$ v/s $\rm{kT_{e}}$}
In Fig. \ref{figure-16} the correlation between $\Gamma$ and $\rm{kT_{e}}$ is shown. The orange and black dashed lines represent the ordinary and simulated least square fit to the data. The correlation between $\Gamma$ and $\rm{kT_{e}}$ is found to be not significant.

\subsubsection{$R$ v/s $\rm{kT_{e}}$}
The correlation between the distant reflection fraction $R$ and $\rm{kT_{e}}$ is presented in the bottom panel of Fig. $\ref{figure-16}$. We did not find any correlation between these two parameters. 

\subsection{Nature of corona in AGN}
\begin{figure}
\vbox{
      \includegraphics[scale=0.54]{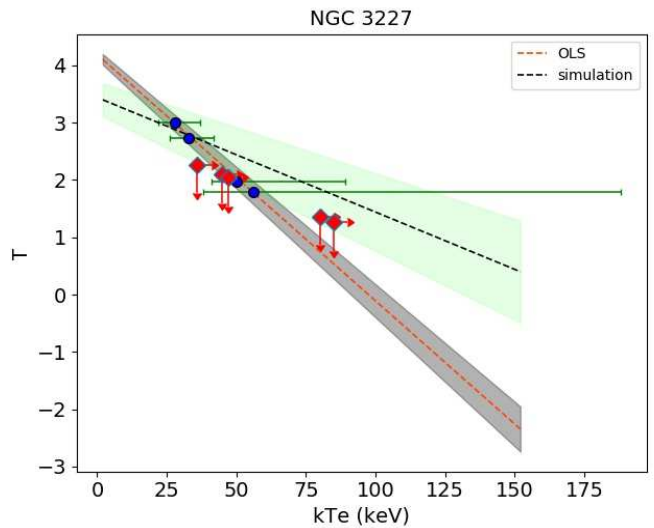}
      }
\caption{Correlation between $\tau$ and $\rm{kT_{e}}$  of 
	NGC 3227. The dashed lines and the shaded regions have the same meaning as in Fig. \ref{figure-13}. Points shown as red in the Figure were not considered for the linear least squares fit.} \label{figure-18}
\end{figure}

The primary X-ray emission from thermal Comptonization 
depends on the optical depth $\tau$ and $\rm{kT_{e}}$ as 
\citep{1996MNRAS.283..193Z,1999MNRAS.309..561Z},
\begin{equation}
    \tau = \sqrt{\frac{9}{4} + \frac{3}{\theta\Big[\Big(\Gamma + \frac{1}{2}\Big)^2 - \frac{9}{4}\Big]}} - \frac{3}{2}  \\
\end{equation}   
where, $\theta = {kT_e}/{m_{e}c^2}$. We show in Fig. \ref{figure-18} the variation 
of the reflection fraction, the optical  depth and the Compton \say{$y$} parameter
with flux. The Compton $y$ parameter is defined 
as \citep{2001ApJ...556..716P},
\begin{equation}
    y \simeq 4\Bigg(\frac{4kT_e}{mc^2}\Bigg)\Bigg[1+\Bigg(\frac{4kT_e}{mc^2}\Bigg)\Bigg] \tau(1+\tau)
\end{equation}

\begin{figure}
\vbox{
      \includegraphics[scale=0.50]{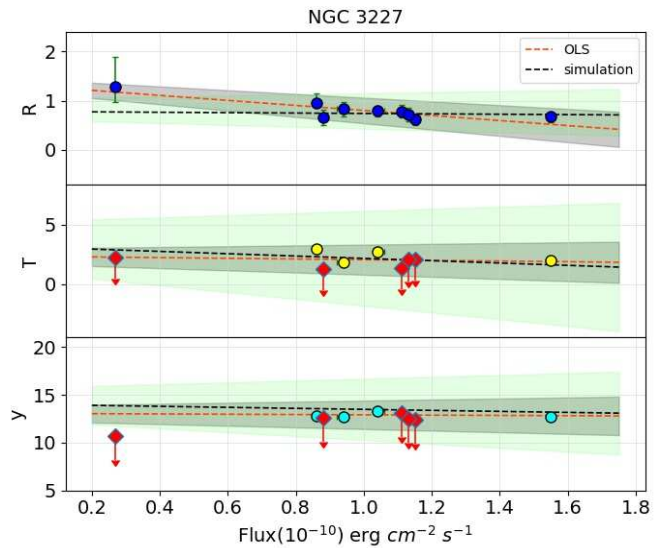}
      }
\caption{Distant reflection fraction $R$, the optical 
	depth $\tau$ and Compton-y parameter as a function of flux for the sources NGC 3227. The dashed lines and the shaded regions have the same meaning as in Fig. \ref{figure-13}. Points shown as red in the Figure were not included in the linear least squares fit.} \label{figure-17}
\end{figure}

According to \cite{1995ApJ...449L..13S}, a Comptonized corona must have a constant 
$y$ in equilibrium. We too found no correlation of $y$ with the flux of the source. The parameter $\tau$ is found not to show any statistically significant variation with flux. $R$ is found to be not correlated with flux(see Fig. \ref{figure-17}).

Multiple epochs of observations on few AGN available today point to increased 
$\rm{E_{cut}}$ with the flux of the sources.  For example, 
from analysis of {\it NuSTAR} and XMM-Newton observations of NGC 5548, 
\cite{2018ApJ...863...71Z} found $\rm{E_{cut}}$ to be positively correlated 
with the flux of the source.  Similarly in Mrk 335 too, \cite{2016MNRAS.456.2722K} 
through joint fitting of {\it Suzaku} and {\it NuSTAR} found $\rm{E_{cut}}$ to be positively 
correlated with flux.  Recently from flux resolved spectroscopy of Ark 564, 
\cite{2020MNRAS.492.3041B} found the temperature of the corona to decrease with 
increasing flux. Also, for the source ESO 103$-$035 \cite{2021ApJ...921...46B} found a positive correlation between $\rm{kT_{e}}$ and flux. Scenarios that could cause the change in the temperature of the corona or $\rm{E_{cut}}$ in
AGN are (a) Compton cooling and (b) expanding corona. In the Compton
cooling scenario, as the source brightens, there will be increased 
seed UV photons from the disk for Comptonization leading to a cooling
of the corona. This leads to a 
\say{cooler when brighter behaviour}. 
In the 
expanding corona scenario, the increase in $\rm{kT_{e}}$ during high flux 
states of the AGN could be due to changes in the geometry of the corona.
This has been invoked to explain the $\rm{E_{cut}}$ changes in 
Mrk 335 \citep{2016MNRAS.456.2722K} and NGC 5548 \citep{2018ApJ...863...71Z} .
According to the expanding corona model, at low flux state,
the corona is warm, optically thick, compact and close to the black hole. This
causes more illumination of the disk leading to larger reflection
fraction. As the source increases in 
brightness, the corona
expands, the optical depth drops and the temperature rises. A reduced
reflection fraction during this period argues that the corona expands 
vertically from the disk. In NGC 3227 too, the corona temperature is 
found to vary with time. We found a negative correlation between the optical depth $\tau$ and $\rm{kT_{e}}$, with $\tau$ decreasing towards higher temperatures (see Fig. \ref{figure-18}). We calculated $\tau$ using Equation 7 and according to \cite{1996MNRAS.283..193Z} $\tau$ is geometry dependent and equals the radial optical depth in an uniform sphere. The negative correlation between $\tau$ and $\rm{kT_{e}}$ argues for a change in
the geometry of the corona \citep{2014ApJ...794...62B}. \cite{2018A&A...614A..37T} too from the analysis of a sample of  AGN found a negative correlation between $\tau$ and $\rm{kT_{e}}$. According to the authors this negative correlation could not be explained with a fixed disk-corona configuration in radiative balance. The possible explanation for this kind of behaviours could be (a) due to the change in the geometry and position of the corona and/or (b) variation in the fraction of the intrinsic disk emission to the total disk emission. We note that broad band spectral energy distribution fits to simultaneous UV to hard X-ray data alone will be able to provide strong constrain on $\tau$. This in turn can put constrains on the role of accretion disk emission to $\rm{kT_{e}}$ changes. In NGC 3227 we found no statistically significant correlation between $\rm{kT_{e}}$ and flux, $R$ and flux, $\Gamma$ and flux as well as $\Gamma$ and $\rm{kT_{e}}$. 

\section{Summary}
In this work, we carried out spectral analysis of the {\it NuSTAR} data 
for three Seyfert type AGN namely NGC 3227, NGC 5548 and MR 2251$-$178 including a few OBSIDs not analyzed yet in comparison to previous works. We summarize our results below

\begin{enumerate}
\item All the sources were found to show moderate variations in their average brightness during the epochs analyzed in this work.
\item In NGC 3227, we found unambiguous evidence for the change in the 
temperature of the corona. This change in $\rm{kT_{e}}$ is also reflected in the variation in $\rm{E_{cut}}$.
For NGC 5548 and MR 2251$-$178 we found no evidence for the variation in the temperature of the corona.
\end{enumerate}

Our knowledge on the variation in the temperature of the corona is known only for less than half a dozen sources. Details on such coronal temperature variation in more AGN are needed to pin point the reasons for the temperature of the corona to vary and its effect on the other physical properties of the sources.

\section{Acknowledgements}
We thank the anonymous referee for her/his useful comments and suggestions which improved the quality and the clarity of the paper. We thank the {\it NuSTAR} Operations, Software and Calibration teams for support with the execution and analysis of these observations. This research has made use of the {\it NuSTAR} Data Analysis Software (NuSTARDAS) jointly developed by the ASI Science Data Center (ASDC, Italy) and the California Institute of Technology (USA). This research has made use of data and/or software provided by 
the High Energy Astrophysics Science Archive Research Center (HEASARC), which is a service of the Astrophysics Science Division at NASA/GSFC.

\bibliographystyle{aa} 
\bibliography{example} 

\label{lastpage}
\end{document}